\documentclass[aps,prl,reprint,amsmath,amssymb,superscriptaddress]{revtex4-2}

\usepackage{graphicx}
\usepackage{bm}
\usepackage{hyperref}
\usepackage{natbib}
\usepackage{lineno}

\begin{document}

\title{Topological Surface Charge Detection via Active Capacitive Compensation: A Pathway to the 4D Quantum Hall Effect}

\author{Yuanze Li}
\affiliation{State Key Laboratory of Low Dimensional Quantum Physics, Department of Physics, Tsinghua University, Beijing 100084, China}

\author{Renfei Wang}
\affiliation{International Center for Quantum Materials, Peking University, Beijing 100871, China}

\author{Yifan Zhang}
\affiliation{School of Information Science and Technology, ShanghaiTech University, Shanghai 201210, China}

\author{Jiahao Chen}
\affiliation{State Key Laboratory of Low Dimensional Quantum Physics, Department of Physics, Tsinghua University, Beijing 100084, China}

\author{Yingdong Deng}
\affiliation{School of Physical Science and Technology, ShanghaiTech University, Shanghai 201210, China}

\author{Jin Xie}
\affiliation{School of Physical Science and Technology, ShanghaiTech University, Shanghai 201210, China}

\author{Xufeng Kou}
\affiliation{School of Information Science and Technology, ShanghaiTech University, Shanghai 201210, China}
\affiliation{ShanghaiTech Laboratory for Topological Physics, School of Physical Science and Technology, ShanghaiTech University, Shanghai 201210, China}

\author{Yang Liu}
\affiliation{International Center for Quantum Materials, Peking University, Beijing 100871, China}

\author{Tian Liang}
\email{tliang@mail.tsinghua.edu.cn}
\affiliation{State Key Laboratory of Low Dimensional Quantum Physics, Department of Physics, Tsinghua University, Beijing 100084, China}
\affiliation{Frontier Science Center for Quantum Information, Beijing 100084, China}

\begin{abstract}

The topological magnetoelectric effect (TME) in three-dimensional topological insulators (TIs), described by $\Delta P = \frac{e^2}{2h} N_{\rm Ch}^{(2)} \Delta B$, serves as a condensed-matter realization of the four-dimensional quantum Hall effect (4D QHE). In dual-gate axion-insulator devices, the TME-induced polarization yields a current $I_{\rm TME} \propto (C_{\rm total}/C_{\rm S})\,Q_{\rm 4D\mathrm{-}QHE}$, where the signal is suppressed by the capacitance ratio $C_{\rm total}/C_{\rm S}$. Here we propose an active compensation scheme that introduces a tunable negative capacitance $C_{\rm comp} \approx -C_{\rm gate}$ into the gate line, effectively canceling the gate dielectric capacitance and driving $C_{\rm total}/C_{\rm S} \to 1$. We validate the method using a quantum anomalous Hall (QAH) device, which shares the same surface-state physics as the axion insulator but permits direct charge measurement via a single gate, recovering over $95\%$ of the quantized charge signal from an initially half-attenuated state. This compensation method provides a robust means of resolving minute TME signals, offering a promising pathway toward direct measurements of the 4D QHE.

\end{abstract}

\maketitle

\subsection{Introduction}

The topological magnetoelectric effect (TME) in three-dimensional topological insulators (3D TIs) represents a condensed-matter realization of the four-dimensional quantum Hall effect (4D QHE), in which the four-dimensional parameter space comprises the three spatial dimensions together with time \cite{Zhang2001,Qi2008,Essin2009}. The TME is characterized by the relation
\begin{equation}
\Delta P = \frac{e^2}{2h} N_{\mathrm{Ch}}^{(2)} \Delta B,
\end{equation}
where $N_{\mathrm{Ch}}^{(2)}$ is the second Chern number (an integer), $\Delta B$ denotes the variation in the applied magnetic field, and $\Delta P$ represents the corresponding change in electric polarization. The TME manifests in a 3D TI when its surface magnetizations are uniformly aligned either all inward or all outward, forming an axion insulator (AI) state \cite{Mogi2017,Mogi2017a,Xiao2018,Liu2020,XuYang2014,Zhuo2023,Bai2024}. In this state, $\Delta B$ induces a quantized parallel $\Delta P$ in the bulk, governed by the coefficient $\frac{e^2}{2h} N_{\mathrm{Ch}}^{(2)}$. In a thin AI film subjected to a perpendicular magnetic field, this polarization change leads to quantized surface charge accumulation of opposite signs on the top and bottom surfaces ($\frac{e^2}{2h}\Delta B$ on one surface and $-\frac{e^2}{2h}\Delta B$ on the other). This accumulation is directly measurable as an out-of-plane current in dual-gate AI devices,
\begin{equation}
I_{\mathrm{TME}} \propto \frac{C_{\mathrm{total}}}{C_{\mathrm{S}}} Q_\mathrm{4D-QHE},
\end{equation}
where $Q_\mathrm{4D-QHE} = A_\mathrm{S} \Delta B \cdot \frac{e^2}{2h} N_{\mathrm{Ch}}^{(2)}$ is the quantized polarization charge originating from the 4D QHE ($A_\mathrm{S}$ is the sample area), $C_{\mathrm{S}}$ is the geometric capacitance between the top and bottom surfaces of the TI film, and $C_{\mathrm{total}} = (1/C_\mathrm{S}+1/C_\mathrm{gate})^{-1}$ is the total device capacitance, consisting of $C_{\mathrm{S}}$ in series with the gate dielectric capacitances $C_\mathrm{gate}$.

Although $Q_{\mathrm{4D-QHE}}/\Delta B$ itself is sufficiently large (approximately $2\,\mathrm{fC/Gs}$ per 1 mm$^2$ sample area) to be detected by ultra-sensitive out-of-plane capacitive techniques \cite{Li2025}—an approach accessible to both the 4D QHE and 2D QHE, in sharp contrast to the in-plane charge pumping method \cite{1990_Russia_origin,1992_Russia_GaAs_chargepumping_Streda,RIKEN_CBST_chargepumping} which is applicable only to the 2D QHE—the actually measurable TME signal is reduced by the geometric-capacitance factor
\begin{equation}
\gamma_{\mathrm{geo}}
=
\frac{C_{\mathrm{total}}}{C_{\mathrm{S}}}
\approx
\frac{C_\mathrm{gate}}{C_{\mathrm{S}}}.
\end{equation}
Therefore, direct detection of the 4D-QHE response requires not only ultra-sensitive charge detection, but also a sufficiently large $\gamma_{\mathrm{geo}}$ so that the TME charge is not attenuated below the detector resolution (approximately $0.1\,\mathrm{fC/Gs}$). 

In surface charge measurements, the gate capacitance also controls the charge-relaxation time (see Supplemental Material~\cite{SuppMat}, Sec. V for details),
\nocite{inhofer2017,inhofer2018,dartiailh2020}
\begin{equation}
\tau_{\mathrm{decay}} \approx \frac{C_\mathrm{gate}}{\sigma_{xx}} .
\end{equation}
Thus, increasing $C_\mathrm{gate}$ simultaneously enhances the measurable TME signal and suppresses dissipative relaxation. For conventional few-nanometer-thick AI films, $\gamma_{\mathrm{geo}}$ is typically only on the order of $10^{-3}$, making the TME signal far below the resolution, even without considering the decay effect quantified by $\tau_{\mathrm{decay}}$. Enlarging the AI sample thickness can increase $\gamma_{\mathrm{geo}}$; however, even for the reported $\sim 100$ nm AI device \cite{Zhuo2023}, an estimate using a relative dielectric constant $\epsilon_{\mathrm{AI}}\sim100$ gives only $\gamma_{\mathrm{geo}}\approx0.03$, corresponding to a TME signal barely comparable to or below the current noise floor. Moreover, the finite longitudinal conductivity of this device, $\sigma_{xx}\sim 1\times10^{-6}$ S, gives a charge-relaxation rate $1/\tau_{\mathrm{decay}}$ of order kHz, whereas a low-crosstalk field-induced charge measurement is limited to a lower frequency range of several hundred Hz or below \cite{Li2025}.

To overcome these limitations, we introduce active capacitive compensation as an enabling capacitance-engineering technique for direct TME measurements. This method directly enhances both the geometric factor $\gamma_{\mathrm{geo}}$ and the charge-relaxation time $\tau_{\mathrm{decay}}$ by increasing the effective gate capacitance, without requiring further improvements in AI-film growth or device geometry. By introducing an effective negative capacitance $C_{\mathrm{comp}} \equiv -C_1$ in series with the gate line, the capacitive compensation alters the effective gate capacitance as
\begin{equation}
C_\mathrm{gate}^{\mathrm{eff}} = \left( \frac{1}{C_\mathrm{gate}}
- \frac{1}{C_1} \right)^{-1},
\end{equation} 
which exceeds the original $C_\mathrm{gate}$. By tuning $C_1$ close to $C_\mathrm{gate}$, this method can substantially enhance $\gamma_{\mathrm{geo}}$ and $\tau_{\mathrm{decay}}$, making direct 4D-QHE detection feasible not only in rare hundred-nanometer-thick AI films, but also in more typical few-nanometer AI films (see below for details). Because a device platform combining a high-quality AI state with a sufficiently large $\gamma_{\mathrm{geo}}$ without compensation has not yet been established, active capacitive compensation constitutes an indispensable ingredient for direct TME measurements.

Below, we experimentally validate this approach using a TI sample in the quantum anomalous Hall (QAH) state \cite{yu2010QAH,kane2010TI,SCZhang2011,ChangCZ2013_QAH_CBST,KouXufeng2014_QAH_CBST,Tokura2014_QAH_CBST,ZhangYuanbo2020_QAH_MBT}, which shares the same surface-state physics as the axion insulator—including identical attenuation mechanisms—but exhibits a field-induced net charge accumulation measurable via a single gate, thereby offering a convenient platform for method development. For the QAH sample, the field-induced quantized charge accumulation $Q_\mathrm{QAH} = \pm A_\mathrm{S} \Delta B \cdot \frac{e^2}{h}$ induces an image charge of equal magnitude and opposite sign on the gate, enabling direct measurement of $Q_{\mathrm{QAH}}$ via the gate readout. By tuning the expected effective gate capacitance through active capacitive compensation, we systematically modify the $\sigma_{xx}$-induced attenuation of this charge signal, quantitatively verifying the compensation control of the effective gate capacitance. Ultimately, using this compensation method, we recover the quantized charge accumulation from an attenuated response.

\subsection{Active capacitive compensation for surface charge measurements}

\begin{figure*}[htbp]
  \centering
  \includegraphics[width=1\linewidth]{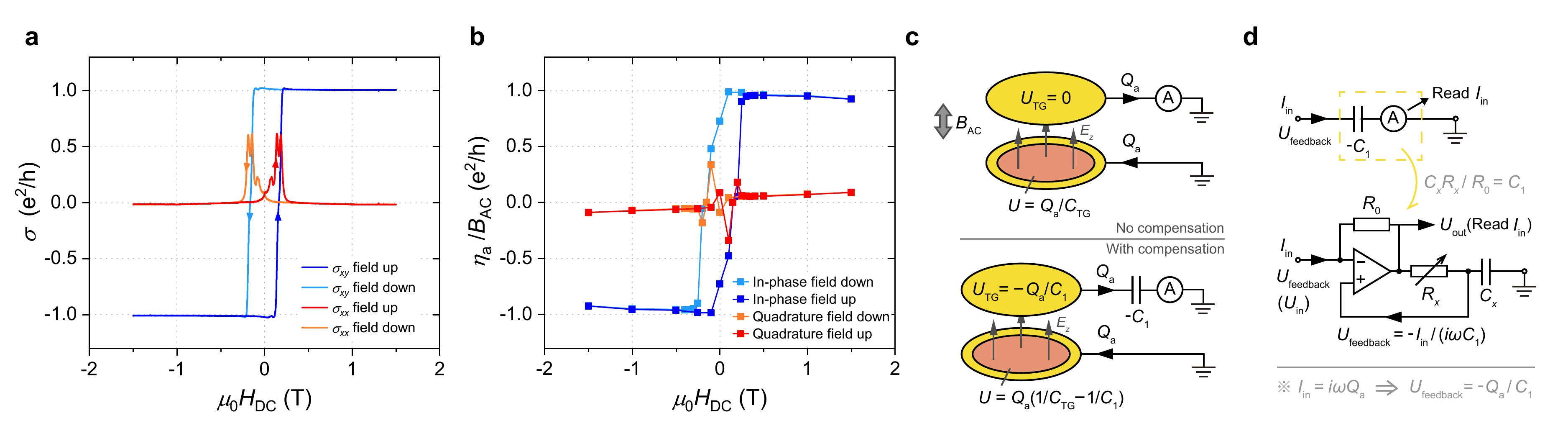}
  \caption{\textbf{Active capacitive compensation and QAH charge accumulation.}
  (a) DC magnetic-field dependence of $\sigma_{xx}$ and $\sigma_{xy}$ of the QAH sample measured on a Hall bar device.
  (b) Field-induced surface charge accumulation $\eta_\mathrm{a}/B_\mathrm{AC}$ measured on the simple disk-shaped sample without compensation. The measurement was performed in the ultra-low-$\sigma_{xx}$ regime at $f = 277.777$ Hz. In-phase and quadrature components are shown separately.
  (c) Schematics of uncompensated and compensated charge-accumulation measurements. In the compensated case, the feedback voltage $U_\mathrm{TG}=-Q_\mathrm{a}/C_1$ suppresses the sample potential buildup.
  (d) Circuit realization of the feedback negative capacitance, with $C_1 = C_x R_x / R_0$. The accumulated charge is obtained from the output voltage as $Q_\mathrm{a} = -U_\mathrm{out}/(i\omega R_0 + 1/C_1)$.}
  \label{fig:fig1}
\end{figure*}

A six-quintuple-layer (6-QL) chromium-doped $\mathrm{(Bi,Sb)}_2\mathrm{Te}_3$ film grown by molecular beam epitaxy (MBE) \cite{KouXufeng2022-CBST-thickness} was fabricated into devices as the QAH platform for the charge accumulation measurement. The film was patterned into Hall bars for transport measurements and disk structures (simple disks and Corbino disks) for charge signals. The Hall bar sample exhibited quantized Hall conductivity ($\sigma_{xy} = \pm e^2/h$) and vanishing longitudinal conductivity ($\sigma_{xx}$) with hysteresis loops, measured at $T<50$ mK [Fig.~\ref{fig:fig1}(a)]. For a more accurate $\sigma_{xx}$, we conducted a two-terminal measurement in the Corbino disk-shaped sample, yielding $\sigma_{xx} \approx 1 \times 10^{-10}$ S within QAH plateaus at the base temperature ($T \approx 10$ mK) of the dilution refrigerator (see Supplemental Material~\cite{SuppMat}, Fig. S2 in Sec. VI).

A uniform alternating magnetic field ($B_\mathrm{AC}$) applied perpendicular to the simple disk-shaped sample generated an intrinsically quantized charge accumulation density $\eta_\mathrm{a}^{(0)} = \sigma_{xy} B_\mathrm{AC}$. The extremely low $\sigma_{xx}$ enabled direct quantization at high frequencies. We first measured the directly quantized charge accumulation at the base temperature. We then heated the sample to increase $\sigma_{xx}$, thereby generating signal attenuation that simulates the TME situation and allows verification of the capacitive compensation method (see later sections). Direct measurement detected the accumulated charge ($Q_\mathrm{a} = A_\mathrm{S}\eta_\mathrm{a}$; $\eta_\mathrm{a}$ is the average charge density) via a top gate connected to a current preamplifier [Fig.~\ref{fig:fig1}(c), top] \cite{Li2025}. At $f = 277.777$ Hz, the in-phase component of $\eta_\mathrm{a}/B_\mathrm{AC}$ showed direct quantization, matching the $\sigma_{xy}$ hysteresis, while the quadrature component was negligible [Fig.~\ref{fig:fig1}(b)].

Signal attenuation of the QAH charge accumulation arises from the potential gradient in the sample, which causes dissipation. The top gate is held at zero potential, but the accumulated charge creates a vertical electric field, inducing a potential rise $U = Q_\mathrm{a}/C_\mathrm{TG}$ in the sample ($C_\mathrm{TG}$ is the top gate capacitance, including the geometric capacitance of the dielectric and the quantum capacitance of the sample; see Supplemental Material~\cite{SuppMat}, Sec. II for details) relative to the grounded edge, which drives dissipative currents. Our compensation method uses the same gate measurement but adds an effective series negative capacitance ($-C_1$) in the gate line [Fig.~\ref{fig:fig1}(c), bottom]. This feeds back a potential $U_\mathrm{TG} = -Q_\mathrm{a}/C_1$, offsetting the sample's potential rise. We define the compensation ratio,
\begin{equation}
   \alpha \equiv  C_\mathrm{TG}/C_1 \quad (0 \le \alpha \le 1),
\end{equation}
quantifying the attenuation suppression. In the ideal dissipation-free limit, this creates an effective gate capacitance $C_\mathrm{TG}^\mathrm{eff} = C_\mathrm{TG}/(1-\alpha)$, effectively reducing the gate dielectric thickness. For finite $\sigma_{xx}$, quantitative modeling shows the compensated charge is $Q_\mathrm{a}^\mathrm{comp}=[(1-\gamma)/(1-\alpha\gamma)]Q_\mathrm{a}^{(0)}$, where $\gamma$ characterizes the raw attenuation (see Supplemental Material~\cite{SuppMat}, Sec. II). Full compensation ($\alpha=1$) cancels potential buildup and completely suppresses attenuation ($Q_\mathrm{a}^\mathrm{comp}=Q_\mathrm{a}^{(0)}$). The parasitic capacitance in the gate line affects the charge signal measurement and voltage feedback. However, its effect is removable by adjusting $C_1$ accordingly (see Supplemental Material~\cite{SuppMat}, Sec. XI).

The effective series negative capacitance is achieved by the active compensation circuit, modified from a transimpedance current preamplifier [Fig.~\ref{fig:fig1}(d)]. The circuit measures the current signal and simultaneously generates a feedback voltage ($U_\mathrm{feedback} = - I_\mathrm{in} / (i \omega C_1)$) controlled by elements $R_0$, $R_x$, and $C_x$, where $C_1 = C_x R_x / R_0$. See Supplemental Material~\cite{SuppMat}, Sec. I.C for details of the circuit analysis.

\subsection{Systematic verification of the capacitive compensation method}

\begin{figure*}[htbp]
  \centering
  \includegraphics[width=1\linewidth]{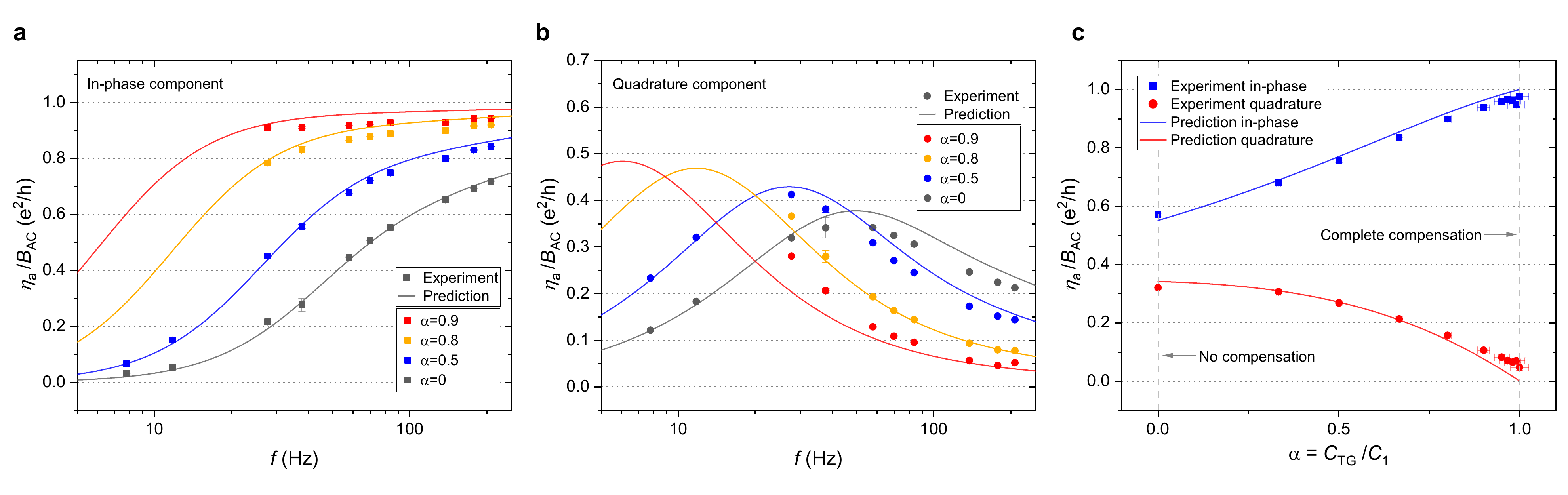}
  \caption{\textbf{Quantitative verification of the capacitive compensation method in QAH charge accumulation.} Measurements were performed on the simple disk-shaped sample at $\mu_0 H_{\mathrm{DC}} = 0.5$ T, with $\sigma_{xx} = 2.35 \times 10^{-8}$ S kept constant throughout the $f$ and $\alpha$ sweeps. Symbols are experimental data and solid lines are theoretical predictions calculated from the dissipation model described in Supplemental Material~\cite{SuppMat}, Sec. II.
  (a) Frequency dependence of the in-phase component of $\eta_\mathrm{a} / B_{\mathrm{AC}}$ for several compensation ratios $\alpha$.
  (b) Frequency dependence of the quadrature component of $\eta_\mathrm{a} / B_{\mathrm{AC}}$ for several $\alpha$.
  (c) In-phase and quadrature components of $\eta_\mathrm{a} / B_{\mathrm{AC}}$ as a function of $\alpha$ at $f = 83.7777$ Hz. Horizontal error bars represent the estimated uncertainty in $\alpha$ arising from systematic errors in circuit elements.}
  \label{fig:fig2}
\end{figure*}

We systematically verified the capacitive compensation method by measuring the frequency ($f$) and compensation ratio ($\alpha$) dependence of charge signals in the disk-shaped QAH sample. To obtain a wider verification range, dissipation was intentionally increased while keeping $\sigma_{xy}$ quantization intact. This was achieved by applying a larger $B_\mathrm{AC}$, which generated mild eddy current heating ($T < 100$ mK). The temperature—corresponding to $\sigma_{xx}=2.35\times 10^{-8}$ S at $\mu_0 H_\mathrm{DC}=\pm 0.5$ T, calibrated from the impedance measurement between the gate and the contact (see Supplemental Material~\cite{SuppMat}, Sec. VIII)—was controlled to remain stable during measurements.

Figures~\ref{fig:fig2}(a) and \ref{fig:fig2}(b) show $\eta_\mathrm{a}/B_\mathrm{AC}$ versus $f$ for $\alpha=0,\,0.5,\,0.8,$ and $0.9$. Experimental data and theoretical simulations calculated from the quantitative dissipation model (see Supplemental Material~\cite{SuppMat}, Sec. II) are compared. In-phase data agree well with theoretical predictions: amplitude grows with increasing $f$ and $\alpha$, finally saturating to the quantized value at high $f$ and large $\alpha$. Quadrature data also follow predicted trends, with peaks shifting to lower frequencies as $\alpha$ increases, reflecting suppressed relaxation. At high $f$ and large $\alpha$, quadrature signals approach zero. A small deviation in the quadrature component observed at low $f$ and large $\alpha$ is primarily attributed to systematic errors in $\alpha$ (e.g., inaccuracies of circuit elements). The $f$ and $\alpha$ dependence of the data can be qualitatively explained by the increased RC time constant $\tau_\mathrm{QAH} = C_\mathrm{TG}^\mathrm{eff}/\sigma_{xx} = C_\mathrm{TG}/\sigma_{xx}/(1-\alpha)$ for larger $\alpha$. Consistent $f$- and $\alpha$-dependent results were obtained at $\mu_0 H_\mathrm{DC} = 0$ T ($\sigma_{xx} = 4.08 \times 10^{-8}$ S; see Supplemental Material~\cite{SuppMat}, Sec. VIII) and in the Corbino disk-shaped sample (see Supplemental Material~\cite{SuppMat}, Sec. X for additional data). Sweeping $\alpha$ at fixed $f=83.7777$ Hz [Fig.~\ref{fig:fig2}(c)] shows that the in-phase component increases toward quantization as $\alpha \to 1$, while the quadrature component is suppressed toward zero. The compensation restores $\eta_\mathrm{a}/B_\mathrm{AC}$ to $(0.97+0.04\,i)\,e^2/h$ at $\alpha=1$ (precision on the order of $1\%$), from $\eta_\mathrm{a}^\mathrm{raw}/B_\mathrm{AC}=(0.55+0.31\,i)\,e^2/h$.

In summary, quantitative control of attenuation via $\alpha$—i.e., tuning $C_\mathrm{TG}^\mathrm{eff}$—is experimentally verified in the measurement of QAH surface charge accumulation, recovering the quantized signal from an initially half-attenuated response.

\subsection{Compensation-enabled quantized QAH charge accumulation}

\begin{figure}[htbp]
  \centering
  \includegraphics[width=0.5\linewidth]{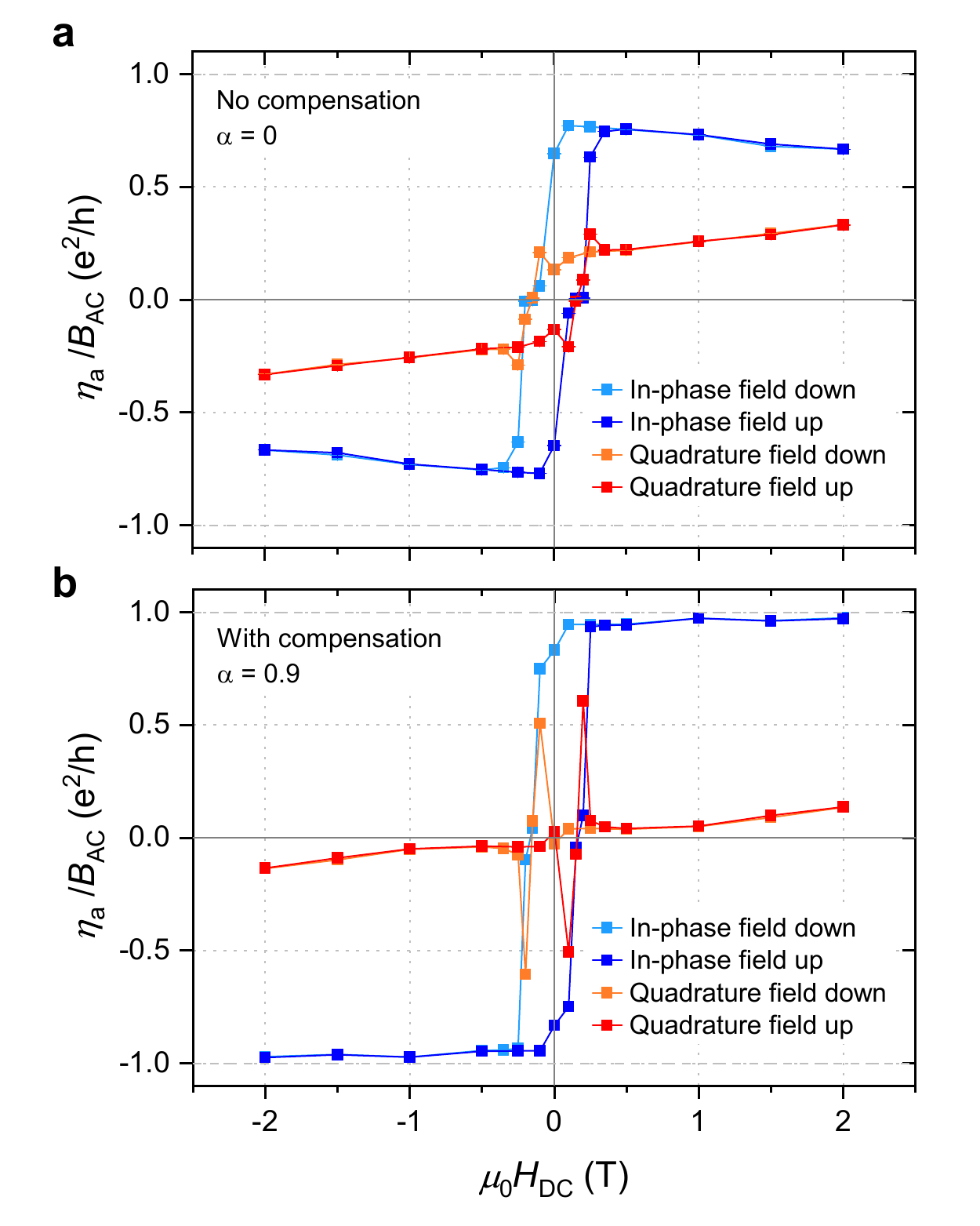}
  \caption{\textbf{Compensation-enabled quantized QAH charge accumulation.} Field dependence of the surface charge accumulation $\eta_\mathrm{a}/B_\mathrm{AC}$ measured on the simple disk-shaped sample at $f = 277.777$ Hz and constant temperature ($\sigma_{xx} =2.35 \times 10^{-8}$ S at $\mu_0 H_{\mathrm{DC}} = 0.5$ T). In-phase and quadrature components are shown separately.
  (a) Uncompensated measurement ($\alpha = 0$), showing strong attenuation and a finite quadrature component.
  (b) Compensated measurement ($\alpha = 0.9$), showing recovered quantization with strongly suppressed quadrature response.}
  \label{fig:fig3}
\end{figure}

Having quantitatively verified the compensation-controlled attenuation suppression in the previous subsection, we performed a full sweep of the DC magnetic field to recover the quantized QAH charge accumulation throughout the entire hysteresis loop. Figure~\ref{fig:fig3} compares the data between uncompensated ($\alpha=0$) and nearly fully compensated ($\alpha=0.9$) cases at $f=277.777$ Hz and a constant temperature ($\sigma_{xx}=2.35\times10^{-8}$ S at $\mu_0 H_{\mathrm{DC}} = 0.5$ T).
Uncompensated signals [Fig.~\ref{fig:fig3}(a)] are clearly attenuated: within QAH plateaus, the in-phase component deviates from quantization, and a finite quadrature component appears. The signals exhibit field-dependent tilt due to the variation of $\sigma_{xx}$. In contrast, with $\alpha=0.9$ [Fig.~\ref{fig:fig3}(b)], signals show nearly flat plateaus: the in-phase component of $\eta_\mathrm{a}/B_\mathrm{AC}$ consistently exceeds $0.95\,e^2/h$, and the quadrature component is reduced to small values. The quadrature-component peaks near the coercive field stem from the surge in $\sigma_{xx}$, where $\alpha=0.9$ is insufficient for full compensation.

The recovered quantized data in Fig.~\ref{fig:fig3}(b) match, and are even better than, the uncompensated signals from the ultra-low-$\sigma_{xx}$ regime [Fig.~\ref{fig:fig1}(b)], despite $\sigma_{xx}$ being two orders of magnitude larger here.
In summary, the capacitive compensation robustly suppresses attenuation across a complete QAH hysteresis loop, recovering quantized charge accumulation plateaus from significantly attenuated raw signals.

\subsection{Discussion of the TME setup}

\begin{figure*}[htbp]
  \centering
  \includegraphics[width=1\linewidth]{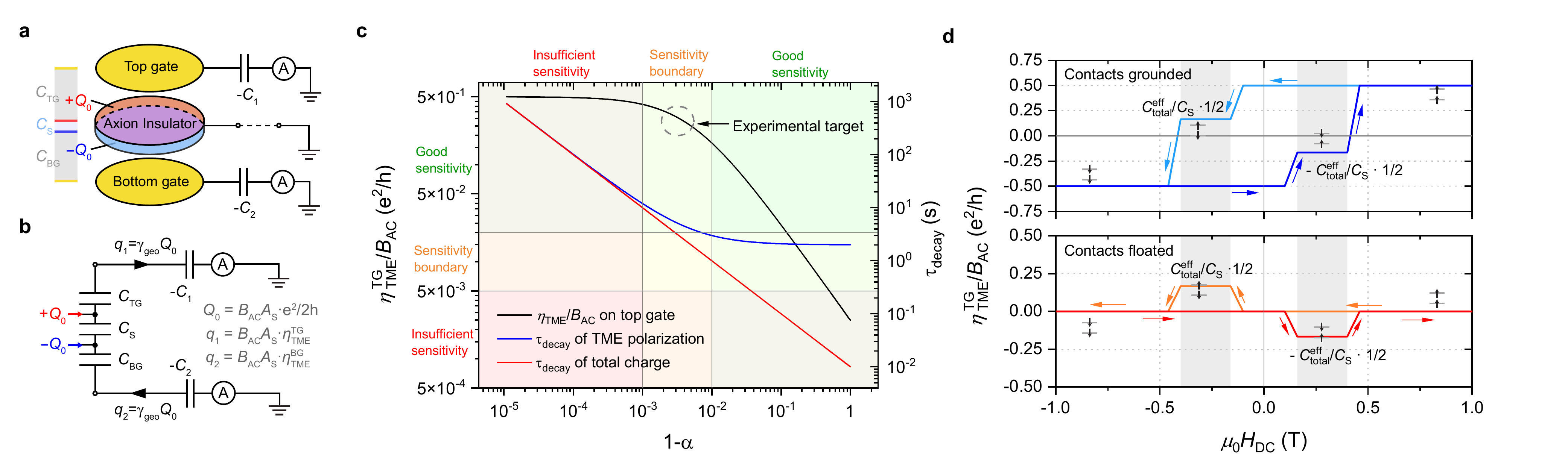}
  \caption{\textbf{Dual-gate capacitive compensation for TME polarization measurements.}
  (a) Schematic of the dual-gate measurement. The compensation generates effective negative capacitances $-C_1$ and $-C_2$ in the gate lines.
  (b) Schematic of the effective circuit. The intrinsic TME surface charges $\pm Q_\mathrm{4D-QHE}$ (abbreviated as $\pm Q_0$ in the figure) induce gate signals $q_1$ and $q_2$. 
  (c) Simulated top-gate TME signal $\eta_\mathrm{TME}^\mathrm{TG}/B_\mathrm{AC}$ and decay times as a function of $(1-\alpha)$, where $\alpha = (1/C_1 + 1/C_2)/(1/C_{\mathrm{TG}} + 1/C_{\mathrm{BG}})$. Background shadings indicate the practical sensitivity intervals, and the dashed circle marks the experimental target region.
  (d) Simulated DC magnetic-field dependence of the compensated $\eta_\mathrm{TME}^\mathrm{TG}/B_\mathrm{AC}$ using the target parameters indicated in (c), for contact-grounded (top) and contact-floating (bottom) configurations. Black arrows indicate the magnetization configurations of the sample surfaces, while blue and red arrows denote the field sweep directions. Gray shaded regions denote the axion-insulator regime.}
  \label{fig:fig4}
\end{figure*}

Since the active capacitive compensation scheme has been quantitatively validated in single-gate measurements, we now analyze its application to the dual-gate geometry proposed for detecting the TME polarization. Figure~\ref{fig:fig4}(a) sketches the setup: two independent gate lines—each identical to that used in the previous single-gate charge measurement—are connected to the top and bottom gates. The geometric capacitances $C_\mathrm{TG}$, $C_\mathrm{BG}$, and $C_\mathrm{S}$ are calibrated via two-terminal measurements and directly determine the compensation precision (see Supplemental Material~\cite{SuppMat}, Sec. IX).

The equivalent circuit in Fig.~\ref{fig:fig4}(b) models the TME polarization $\pm Q_\mathrm{4D-QHE} = \pm B_\mathrm{AC} A_\mathrm{S} \cdot e^2/(2h)$. These induce gate charges $q_1$ and $q_2$ through the capacitance network (see Supplemental Material~\cite{SuppMat}, Sec. IV for the derivation). In the ideal limit of infinitely large gate capacitances ($C_\mathrm{TG}, C_\mathrm{BG} \to \infty$), $q_1 = q_2 = Q_\mathrm{4D-QHE}$. In real devices, however, finite gate capacitances cause attenuation through two mechanisms: dissipation-induced relaxation (characterized by the time constant $\tau_\mathrm{TME}$) and purely electrostatic screening (quantified by the geometric factor $\gamma_\mathrm{geo} = C_\mathrm{total}/C_\mathrm{S}$), with $C_\mathrm{total}$ being the series combination of $C_\mathrm{TG}$, $C_\mathrm{BG}$, and $C_\mathrm{S}$. Notably, $\tau_\mathrm{TME} = \tau_\mathrm{QAH} / \gamma_\mathrm{geo}$; although the TME signal is electrostatically attenuated, the associated dissipation is correspondingly reduced (see Supplemental Material~\cite{SuppMat}, Sec. V).

To mitigate attenuation, we apply independent voltage feedback in each gate line, creating effective negative capacitances $-C_1$ and $-C_2$. This yields enlarged effective gate capacitances $C_\mathrm{TG}^\mathrm{eff} = (1/C_\mathrm{TG} - 1/C_1)^{-1}$ and analogously for $C_\mathrm{BG}^\mathrm{eff}$. We define the dual-gate compensation ratio $\alpha = (1/C_1 + 1/C_2)/(1/C_\mathrm{TG} + 1/C_\mathrm{BG})$; with symmetric compensation, $C_\mathrm{TG}/C_1 = C_\mathrm{BG}/C_2 = \alpha$. Then $\gamma_\mathrm{geo}$ becomes a known function of $\alpha$ and the calibrated capacitances, enabling quantitative recovery of the intrinsic TME signal even when $\alpha \neq 1$.

Perfect compensation ($\alpha = 1$) is experimentally challenging, but partial compensation already significantly reduces attenuation: in prior QAH measurements, $\alpha = 0.9$ recovered over 95\% of the quantized response. For TME measurements, we can similarly choose $\alpha$ sufficiently close to 1 to maximize the signal-to-noise ratio.

Figure~\ref{fig:fig4}(c) simulates the $\alpha$ dependence of the measurable TME signal $\eta_\mathrm{TME}^\mathrm{TG}/B_\mathrm{AC} = \gamma_\mathrm{geo} \cdot e^2/(2h)$ and the decay time $\tau_\mathrm{TME}$, using typical parameters ($C_\mathrm{TG} = C_\mathrm{BG} = 1$ nF, $C_\mathrm{S} = 100$ nF, $\sigma_{xx} = 10^{-7}$ S). As $\alpha \to 1$, the signal approaches $e^2/(2h)$ and $\tau_\mathrm{TME}$ increases. Importantly, the signal remains purely in-phase for any $\alpha$, unlike dissipation-induced quadrature components. Our measurement frequency (up to 500 Hz) far exceeds $1/\tau_\mathrm{TME}$, so dissipation is effectively suppressed; thus, compensation efficacy is primarily governed by $\gamma_\mathrm{geo}$.

We visualize the experimental feasibility using two practical sensitivity boundaries (calculated based on the data uncertainty in QAH charge-accumulation experiments): $(1-\alpha) \sim 10^{-3}$ (compensation-setting limit) and $\eta_\mathrm{TME}^\mathrm{TG}/B_\mathrm{AC} \sim 5\times10^{-3}\, e^2/h$ (signal floor for $\sim 5$ mm$^2$ area). At $\alpha = 0$, the signal lies below this floor, but intermediate $\alpha$ values fall within the region where both parameters are measurable with sufficient sensitivity. The dashed circle in Fig.~\ref{fig:fig4}(c) indicates a practical target $\alpha$ range balancing signal strength and compensation accuracy. Crucially, even if the raw signal is only $\sim 50\%$ of the ideal value, $\gamma_\mathrm{geo}$ can be precisely calculated, allowing quantitative extraction of the intrinsic TME signal via $(\eta_\mathrm{TME}^\mathrm{TG}/B_\mathrm{AC})/\gamma_\mathrm{geo}$.

Finally, Figure~\ref{fig:fig4}(d) simulates the compensated TME hysteresis loop using $C_\mathrm{TG}^\mathrm{eff} = C_\mathrm{BG}^\mathrm{eff} = C_\mathrm{S} = 100$ nF. In a magnetic topological insulator with different top/bottom coercive fields, the TME polarization appears in the axion insulator regime (opposite surface magnetizations), while the QAH charge accumulation occurs in aligned magnetization regions. With compensation, the TME signal exhibits quantized plateaus comparable in magnitude to the QAH signal, well above the experimental noise floor. Moreover, because the TME polarization and compensation circuit operate without requiring sample contacts, the contacts can be floated to isolate a clean TME plateau, free from QAH charge-pumping artifacts. 

\subsection{Conclusion}

In conclusion, we have experimentally validated an active capacitive compensation method that enables quantitative recovery of attenuated surface charge signals in topological systems. By introducing an effective negative capacitance to control the measurement bandwidth and dissipation, the method recovered over 95\% of the intrinsically quantized charge accumulation from an initially half-attenuated state in quantum anomalous Hall samples, with experimental data closely matching theoretical predictions.

This compensation scheme is directly applicable to dual-gate TME measurements, where tuning the total device capacitance $C_{\mathrm{total}}$ overcomes the geometric attenuation factor  $\gamma_\mathrm{geo} = C_{\mathrm{total}}/C_{\mathrm{S}}$ that hinders detection. Our results establish a robust pathway toward direct experimental observation of the 4D QHE via TME measurements in axion insulator devices.

\newpage

\noindent\textit{Acknowledgments}---We thank Xing~He and Wanjun~Jiang for providing access to the argon ion etching system used for sample fabrication, and Mengqiao~Geng for technical support with dilution-refrigerator operation. Y.~Li thanks Lewei~Li for assistance with schematic illustrations. This work was supported by the National Key R\&D Program of China (Grants No. 2021YFA1401600 and No. 2021YFA1401900).

\par\medskip
\noindent\textit{Data availability}---The data that support the findings of this article are not publicly available. The data are available from the authors upon reasonable request.

\par\medskip
\noindent\textit{Author contributions}---T.~L. conceived, designed, and supervised the project, and coordinated collaborations among the research groups. Y.~Li designed the capacitive compensation scheme for the ultra-high-sensitivity surface charge detection and performed the measurements with extensive assistance from R.~W., J.~C., and Y.~Liu. The high-quality quantum anomalous Hall thin films were grown by Y.~Z. under the supervision of X.~K. and subsequently fabricated into devices by Y.~Li with support from Y.~D. and J.~X. Y.~Li and T.~L. analyzed the experimental data and calculated the theoretical models of the TME. The manuscript was written by Y.~Li and T.~L. with input from all authors. All authors discussed the results and provided feedback on the manuscript.

\bibliography{MainPaper}

\end{document}


\title{Supplemental Material for ``Topological Surface Charge Detection via Active Capacitive Compensation: A Pathway to the 4D Quantum Hall Effect''}

\author{Yuanze Li}
\affiliation{State Key Laboratory of Low Dimensional Quantum Physics, Department of Physics, Tsinghua University, Beijing 100084, People's Republic of China}

\author{Renfei Wang}
\affiliation{International Center for Quantum Materials, Peking University, Beijing 100871, People's Republic of China}

\author{Yifan Zhang}
\affiliation{School of Information Science and Technology, ShanghaiTech University, Shanghai 201210, People's Republic of China}

\author{Jiahao Chen}
\affiliation{State Key Laboratory of Low Dimensional Quantum Physics, Department of Physics, Tsinghua University, Beijing 100084, People's Republic of China}

\author{Yingdong Deng}
\affiliation{School of Physical Science and Technology, ShanghaiTech University, Shanghai 201210, People's Republic of China}

\author{Jin Xie}
\affiliation{School of Physical Science and Technology, ShanghaiTech University, Shanghai 201210, People's Republic of China}

\author{Xufeng Kou}
\affiliation{School of Information Science and Technology, ShanghaiTech University, Shanghai 201210, People's Republic of China}
\affiliation{ShanghaiTech Laboratory for Topological Physics, School of Physical Science and Technology, ShanghaiTech University, Shanghai 201210, People's Republic of China}

\author{Yang Liu}
\affiliation{International Center for Quantum Materials, Peking University, Beijing 100871, People's Republic of China}

\author{Tian Liang}
\email{tliang@mail.tsinghua.edu.cn}
\affiliation{State Key Laboratory of Low Dimensional Quantum Physics, Department of Physics, Tsinghua University, Beijing 100084, People's Republic of China}
\affiliation{Frontier Science Center for Quantum Information, Beijing 100084, People's Republic of China}

\maketitle
\tableofcontents
\clearpage

\section{Methods and samples}

\subsection{Sample growth and fabrication}

The devices, including the simple disk, Corbino disk, and Hall bar structures, were fabricated from the same wafer of a 6-QL chromium-doped $\text{(Bi,Sb)}_2\text{Te}_3$ film grown by molecular beam epitaxy (MBE) on a GaAs (111)B substrate. First, the film was patterned by argon ion milling. Ti/Au (3 nm/47 nm) contact electrodes were then deposited by thermal evaporation. Next, an approximately 200 nm AlO$_x$ gate dielectric was deposited at room temperature using oxygen plasma-assisted atomic layer deposition (ALD), followed by the evaporation of the top gate. All patterning steps were performed using Molybdenum (Mo) masks to avoid the use of solvents. The dielectric layer is much thicker than those in conventional samples to ensure insulation, as the sample area is large. The simple disk-shaped sample has a film radius of 1200 µm, and its top gate overlaps the contact region by 40 µm in width to ensure complete coverage. The Corbino disk-shaped sample (only used in Supplemental Material) has inner and outer radii of 350 µm and 1000 µm, and a partially covered, concentric top gate with inner and outer radii of 450 µm and 900 µm.

\subsection{AC magnetic field generation}

The AC magnetic field was generated by a solenoidal coil wound on a 2 mm-thick cylindrical copper framework thermally anchored to the 1 K still plate. The framework is electrically grounded to shield against capacitive coupling between the coil and sample wires. The coil is constructed using 0.438 mm-diameter Cu-clad NbTi superconducting wire and has a height of 50 mm, an inner diameter of 48 mm, and an inductance of 180 mH. Samples are positioned at the coil center, where the AC field is uniform and perpendicular to the sample plane. The coil is driven by the differential sine outputs of an SR865A lock-in amplifier (50 $\Omega$ output impedance per terminal), providing zero-mean excitation voltage to minimize crosstalk.

The dependence of the amplitude and phase of the AC field on the excitation voltage was calibrated in situ at each measurement frequency using a GaAs/AlGaAs Hall sensor placed near the sample. A 0.01 T DC background field was applied to ensure a linear response. We applied a DC current to pass through the sensor and measured the AC Hall voltage using a lock-in amplifier referenced in-phase with the excitation voltage. The AC field was extracted via the sensor's Hall coefficient ($k_\text{Hall}=\Delta\rho_{yx}/\Delta B$), measured by conventional transport measurements. Currents of $\pm 1$ µA were applied for the calibration, and the results were anti-symmetrized to eliminate background crosstalk. Calibration results are given in Sec. \ref{sec:coil_calibration}.

\subsection{Charge signal measurement with the compensation}

For charge-signal measurements, the contact and top gate of the disk-shaped sample were connected through low-resistance coaxial cables to home-built transimpedance current preamplifiers based on ADA4530-1 chips, with a transimpedance of $R_0=10$ M$\Omega$. The contact line was held at ground potential by grounding the offset input of its preamplifier. The top gate preamplifier output was sent simultaneously to a lock-in amplifier for signal readout and to an RC divider network that generated the compensation voltage, which was fed back to the top gate line through the offset input of the top gate preamplifier. The compensation mechanism, which realizes an effective negative capacitance $-C_1$, is analyzed below using the circuit in Fig.~1(d) of the main text.

The transimpedance stage of the top gate preamplifier converts the input current into an output voltage,
\begin{equation}
U_\text{out} = U_\text{in} - I_\text{in} R_0.
\end{equation}
The RC divider consists of a variable resistor box $R_x$ (0.1 $\Omega$ setting accuracy) in series with a fixed capacitor $C_x=10$ µF. The feedback voltage ($U_\text{feedback}$) is taken from their junction:
\begin{equation}
U_\text{feedback} = \frac{U_\text{out}}{1+i\omega C_x R_x}.
\end{equation}
Between the RC-divider output and the offset input of the top gate preamplifier, $U_\text{feedback}$ was first processed by an SR560 voltage preamplifier set in high-pass filter mode (0.3 Hz, 6 dB) with a gain of 100, and then reduced by a 1:100 divider, thereby keeping $U_\text{feedback}$ unchanged while suppressing low-frequency noise and avoiding overload of the feedback loop.

Since the operational amplifier enforces $U_\text{in}=U_\text{feedback}$, one obtains
\begin{equation}
U_\text{feedback} = -\frac{I_\text{in}}{i\omega C_1}, \qquad C_1 \equiv \frac{C_x R_x}{R_0},
\end{equation}
showing that the feedback acts as an effective tunable capacitance $C_1$ set by $R_x$.

With $I_\text{in}=i\omega Q_\text{a}$, the measured output becomes
\begin{equation}
U_\text{out} = - I_\text{in}\!\left(R_0 + \frac{1}{i\omega C_1}\right)
             = -Q_\text{a}\!\left(i\omega R_0 + \frac{1}{C_1}\right),
\end{equation}
so that
\begin{equation}
Q_\text{a} = -\frac{U_\text{out}}{i\omega R_0 + 1/C_1}.
\end{equation}

In the measurements, $R_x$ was chosen according to the calibrated capacitances and the target compensation ratio $\alpha$. Grounding the offset input of the top gate preamplifier provided the raw measurement without compensation. The reference phase of all lock-in amplifiers was aligned with the AC coil excitation voltage.

For DC-field-dependent measurements, we used a fixed-point method to maintain a stable low temperature: each field point was held for 600 s for stabilization, followed by 300 s of data acquisition.

\section{Quantitative dissipation model for compensated QAH charge accumulation}
\label{sec:QAH_charge_accumulation}

Throughout the theoretical sections below, all voltages, currents, and charge densities denote complex amplitudes at angular frequency $\omega$. The corresponding physical signal is understood as the real part of the complex amplitude multiplied by $e^{i\omega t}$, consistent with the convention used in the main text.

We begin from the generalized in-plane transport equation, 
\begin{equation}
\begin{pmatrix}
J_x \\
J_y
\end{pmatrix}
=
\begin{pmatrix}
\sigma_{xx} & \sigma_{xy} \\
-\sigma_{xy} & \sigma_{xx}
\end{pmatrix}
\left[
\begin{pmatrix}
E_x \\
E_y
\end{pmatrix}
+
\frac{1}{e}
\begin{pmatrix}
\partial_x \mu \\
\partial_y \mu
\end{pmatrix}
\right],
\label{eq:SM_QAH_sigma_tensor}
\end{equation}
where $\sigma_{xx}$ and $\sigma_{xy}$ are the longitudinal and Hall conductances of the QAH sample, $\mu$ is the chemical potential, $e$ is the elementary charge, and $E$ is the electric field. A time-dependent perpendicular magnetic field $\mathbf B=B_z\hat{\mathbf z}$ generates an in-plane electric field through Faraday's law,
\begin{equation}
-\frac{\partial \mathbf B}{\partial t}=\nabla\times\mathbf E
\quad\Rightarrow\quad
\frac{\partial B_z}{\partial t}
=
\frac{\partial E_x}{\partial y}-\frac{\partial E_y}{\partial x}.
\label{eq:SM_QAH_faraday}
\end{equation}
The induced charge accumulation density $\eta(\mathbf r)$ obeys the continuity equation,
\begin{equation}
\frac{\partial \eta}{\partial t}
=
-\nabla\cdot\mathbf J.
\label{eq:SM_QAH_continuity}
\end{equation}
Combining Eqs. \eqref{eq:SM_QAH_sigma_tensor}--\eqref{eq:SM_QAH_continuity}, we obtain
\begin{equation}
\frac{\partial \eta}{\partial t}
=
\sigma_{xy}\frac{\partial B_z}{\partial t}
-\sigma_{xx}\nabla\cdot\mathbf E -
\frac{\sigma_{xx}}{e}\nabla^2\mu
=
\sigma_{xy}\frac{\partial B_z}{\partial t}
+\sigma_{xx}\nabla^2 U -
\frac{\sigma_{xx}}{e}\nabla^2\mu,
\label{eq:SM_QAH_eta_general}
\end{equation}
where $U(\mathbf{r})$ is the electric potential (voltage) in the sample.
Here we assume spatially uniform $\sigma_{xx}$ and $\sigma_{xy}$, so that the transverse component of the chemical-potential-gradient current is divergence-free.

Note that the chemical potential $\mu$ is related to the charge density $\eta$ through the Fermi level variation or quantum capacitance (unit area),
\begin{equation}
    C_\text{q(unit)} = e^2 \frac{\partial n}{\partial \mu} = -e \frac{\partial \eta}{\partial \mu}.
\end{equation}
Therefore Eq.~\eqref{eq:SM_QAH_eta_general} can be written as
\begin{equation}
\frac{\partial \eta}{\partial t}
=
\sigma_{xy}\frac{\partial B_z}{\partial t}
+\sigma_{xx}\nabla^2 U + \sigma_{xx}
\frac{1}{C_\text{q(unit)}}\nabla^2\eta.
\label{eq:SM_QAH_eta_general2}
\end{equation}

We now consider the simple disk-shaped sample, for which the top gate fully covers the sample area $A_\text{S}$. Let the top-gate dielectric thickness and dielectric constant be $d_\text{TG}$ and $\varepsilon_\text{TG}$, respectively. The local electrostatic relation between the sample potential $U(\mathbf r)$ and the top-gate potential $U_\text{TG}$ is
\begin{equation}
U-U_\text{TG}=\frac{d_\text{TG}}{\varepsilon_\text{TG}}\eta
=\frac{A_\text{S}}{C_\text{TG}^\text{(geo)}}\eta,
\label{eq:SM_QAH_electrostatic}
\end{equation}
where
\[
C_\text{TG}^\text{(geo)}=\frac{\varepsilon_\text{TG}A_\text{S}}{d_\text{TG}}.
\]

\subsection{Uncompensated case}

In the uncompensated case, $U_\text{TG}=0$, so
\[
U=\frac{A_\text{S}}{C_\text{TG}^\text{(geo)}}\eta.
\]
For a harmonic magnetic drive $B_z=B_\text{AC}e^{i\omega t}$, Eq. \eqref{eq:SM_QAH_eta_general2} becomes
\begin{equation}
\left(1+\frac{1}{k^2}\nabla^2\right)\eta=\sigma_{xy}B_\text{AC},
\qquad
k^2=\frac{\omega (1/C_\text{TG}^\text{(geo)}+1/C_\text{q})^{-1}}{i\sigma_{xx}A_\text{S}} = \frac{\omega C_\text{TG}}{i\sigma_{xx}A_\text{S}},
\label{eq:SM_QAH_pde}
\end{equation}
where $C_\text{q} = C_\text{q(unit)} A_\text{S}$ is the total quantum capacitance of the sample. We use the total gate capacitance $C_\text{TG}$ to combine the effects of quantum capacitance and geometric capacitance,
\begin{equation}
    C_\text{TG} = (\frac{1}{C_\text{TG}^\text{(geo)}}+\frac{1}{C_\text{q}})^{-1}.
    \label{eq:SM_total_capacitance}
\end{equation}

For a simple disk-shaped sample of radius $r_0$, the finite radial solution is
\[
\eta(r)=\sigma_{xy}B_\text{AC}+a\,J_0(kr),
\]
where $J_0$ is the zeroth-order Bessel function of the first kind. The boundary condition for the sample is the fixed electrochemical potential,
\begin{equation}
    \mu^{*}(r_0) = \mu(r_0) - e U(r_0) = \text{Const.}\,,
\end{equation}
which can be transformed as
\begin{equation}
    \Delta\mu^{*}(r_0) = -e \left[\frac{\eta(r_0)}{C_\text{q(unit)}} +U(r_0)\right] = 0.
    \label{eq:SM_boundary_condition_direct}
\end{equation}
The voltage of the gate ($U_\text{TG}$) determines $\eta(r_0)$ via Eqs. \eqref{eq:SM_boundary_condition_direct} and \eqref{eq:SM_QAH_electrostatic}:
\begin{equation}
    \eta(r_0) = - U_\text{TG} \frac{(1/{C_\text{TG}^\text{(geo)}}+1/{C_\text{q}})^{-1}}{A_\text{S}}
    = - U_\text{TG} \frac{C_\text{TG}}{A_\text{S}}.
    \label{eq:SM_boundary_condition}
\end{equation}
Since the top gate is grounded, Eq. \eqref{eq:SM_boundary_condition} gives $\eta(r_0)=0$, and hence
\begin{equation}
\eta(r)=\sigma_{xy}B_\text{AC}\left[1-\frac{J_0(kr)}{J_0(kr_0)}\right].
\label{eq:SM_QAH_eta_raw}
\end{equation}

The total accumulated charge is
\begin{equation}
Q_\text{a}^{\rm raw}
=
2\pi\int_0^{r_0}r\,\eta(r)\,dr
=
Q_\text{a}^{(0)}(1-\gamma),
\label{eq:SM_QAH_Qraw}
\end{equation}
where
\begin{equation}
Q_\text{a}^{(0)}=A_\text{S}\sigma_{xy}B_\text{AC}
\label{eq:SM_QAH_Q0}
\end{equation}
is the intrinsic QAH charge, and
\begin{equation}
\gamma=\frac{2J_1(kr_0)}{kr_0J_0(kr_0)}
\label{eq:SM_QAH_gamma}
\end{equation}
is the complex attenuation factor, where $J_1$ is the first-order Bessel function of the first kind. Thus $\gamma$ is determined by $\omega$, $\sigma_{xx}$, $C_\text{TG}$, and $A_\text{S}$ (or $r_0$).

\subsection{Compensated case}

We now include the active capacitive compensation. As described in the main text, the compensation circuit is equivalent to a feedback voltage
\begin{equation}
U_\text{TG}=-\frac{Q_\text{a}}{C_1},
\qquad
\alpha\equiv \frac{C_\text{TG}}{C_1},
\label{eq:SM_QAH_feedback}
\end{equation}
where $C_1$ is the feedback capacitance and $\alpha$ is the compensation ratio.

The differential equation remains Eq. \eqref{eq:SM_QAH_pde}; only the boundary condition is modified. Using Eq. \eqref{eq:SM_boundary_condition}, the feedback $U_\text{TG}$ gives
\begin{equation}
\eta(r_0)=\frac{C_\text{TG}}{A_\text{S}C_1}Q_\text{a}
=\frac{\alpha}{A_\text{S}}Q_\text{a}.
\label{eq:SM_QAH_boundary_comp}
\end{equation}
Integrating the corresponding solution over the sample area yields
\begin{equation}
Q_\text{a}
=
Q_\text{a}^{(0)}+\left(\alpha Q_\text{a}-Q_\text{a}^{(0)}\right)\gamma,
\label{eq:SM_QAH_selfconsistent}
\end{equation}
from which
\begin{equation}
Q_\text{a}^{\rm comp}
=
\frac{1-\gamma}{1-\alpha\gamma}\,
Q_\text{a}^{(0)}.
\label{eq:SM_QAH_comp_final}
\end{equation}
Equivalently, the measured average charge density is
\begin{equation}
\eta_\text{a}^{\rm comp}
=
\frac{Q_\text{a}^{\rm comp}}{A_\text{S}}
=
\frac{1-\gamma}{1-\alpha\gamma}\,
\sigma_{xy}B_\text{AC}.
\label{eq:SM_QAH_eta_comp_final}
\end{equation}
This result is the quantitative dissipation model used in the main text [Fig.~2]. In the uncompensated limit $\alpha=0$, it reduces to Eq. \eqref{eq:SM_QAH_Qraw}; in the full-compensation limit $\alpha=1$, one has
\[
Q_\text{a}^{\rm comp}=Q_\text{a}^{(0)},
\qquad
\eta_\text{a}^{\rm comp}=\sigma_{xy}B_\text{AC}.
\]

The corresponding electrochemical driving potential ($V^{*} = \Delta \mu^{*}/(-e)$) is
\begin{equation}
V^{*}(r)
=
\frac{Q_\text{a}^{(0)}}{C_\text{TG}}
\frac{1-\alpha}{1-\alpha\gamma}
\left[
1-\frac{J_0(kr)}{J_0(kr_0)}
\right].
\label{eq:SM_QAH_Ur}
\end{equation}
At $\alpha=1$, $V^{*}(r)=0$ everywhere, i.e. $\mu^{*}$ is spatially uniform, with no charge diffusion current.

\subsection{Generalization to arbitrary sample shape}

The result is not restricted to the disk geometry. For an arbitrary sample shape with uniformly gated area $A_\text{S}$ and grounded sample boundary, the same differential equation,
\[
\left(1+\frac{1}{k^2}\nabla^2\right)\eta=\sigma_{xy}B_\text{AC},
\]
still applies. Writing
\[
\eta(\mathbf r)=\sigma_{xy}B_\text{AC}+\xi(\mathbf r),
\]
one obtains
\[
\left(1+\frac{1}{k^2}\nabla^2\right)\xi=0.
\]
Since the sample edge is grounded, the boundary value of $\eta$ is still consistently $\alpha Q_\text{a}/A_\text{S}$ as in Eq. \eqref{eq:SM_QAH_boundary_comp}. By linearity, the solution can therefore be written as
\[
\xi(\mathbf r)=\left(\frac{\alpha Q_\text{a}}{A_\text{S}}-\sigma_{xy}B_\text{AC}\right)h(\mathbf r),
\]
where $h(\mathbf r)$ is the normalized solution satisfying $h|_{\partial\Omega}=1$. Defining
\begin{equation}
\gamma=\frac{1}{A_\text{S}}\int_{A_\text{S}} h(\mathbf r)\,dS,
\label{eq:SM_QAH_gamma_general}
\end{equation}
one again arrives at Eq. \eqref{eq:SM_QAH_selfconsistent}, and hence
\begin{equation}
Q_\text{a}^{\rm comp}
=
\frac{1-\gamma}{1-\alpha\gamma}\,
Q_\text{a}^{(0)}.
\label{eq:SM_QAH_general_final}
\end{equation}
Therefore, the dependence on $\alpha$ is universal, while the sample geometry enters only through the complex attenuation factor $\gamma$.

\section{Impedance between the top gate and contact in the simple disk-shaped sample}

We now consider the impedance between the top gate and the grounded sample contact, which is used in Sec. \ref{sec:sigma_xx_calibration} for $\sigma_{xx}$ calibration. This problem is governed by the same dissipative electrodynamics as in Sec. \ref{sec:QAH_charge_accumulation}, with the only difference that $B_\text{AC}=0$. The local electrostatic relation remains
\begin{equation}
U-U_t=\frac{d_\text{TG}}{\varepsilon_\text{TG}}\eta
=\frac{A_\text{S}}{C_\text{TG}^\text{(geo)}}\eta,
\label{eq:SM_imp_electrostatic}
\end{equation}
where $U_t$ is the voltage applied to the top gate ($U_\text{TG} = U_t$) and $\eta(\mathbf r)$ is the induced charge density on the sample. The governing equation is directly obtained from Eq. \eqref{eq:SM_QAH_pde}:
\begin{equation}
\nabla^2\eta + k^2\eta=0,
\qquad
k^2=\frac{\omega C_\text{TG}}{i\sigma_{xx}A_\text{S}}.
\label{eq:SM_imp_pde}
\end{equation}

For a circular sample of radius $r_0$, the finite radial solution is
\[
\eta(r)=a\,J_0(kr),
\]
The boundary condition is the same as Eq. \eqref{eq:SM_boundary_condition},
\[
-U_t=\frac{A_\text{S}}{C_\text{TG}}\eta(r_0).
\]
Therefore,
\begin{equation}
\eta(r)=-\frac{C_\text{TG}}{A_\text{S}}U_t\,
\frac{J_0(kr)}{J_0(kr_0)}.
\label{eq:SM_imp_eta}
\end{equation}

Integrating over the sample area, the total charge on the sample is
\begin{equation}
Q_\text{s}
=
- C_\text{TG}U_t\,
\frac{2J_1(kr_0)}{kr_0J_0(kr_0)},
\label{eq:SM_imp_Qs}
\end{equation}
and the top-gate charge is
\begin{equation}
Q_t=-Q_\text{s}.
\label{eq:SM_imp_Qt}
\end{equation}
The measured impedance is then
\begin{equation}
Z_t(\omega)= \frac{U_t}{i\omega Q_t}
=
\frac{1}{i\omega C_\text{TG}}
\frac{kr_0J_0(kr_0)}{2J_1(kr_0)}.
\label{eq:SM_imp_Z}
\end{equation}

The parasitic capacitance between the overlapped top gate and contact, $C_\text{p1}$, contributes an additional parallel branch. The overall measured impedance is therefore
\begin{equation}
Z_\text{TC}
=
\frac{1}{i\omega}
\left[
C_\text{p1}+C_\text{TG}\frac{2J_1(kr_0)}{kr_0J_0(kr_0)}
\right]^{-1}.
\label{eq:overall_impedance}
\end{equation}
For fixed sample geometry, $Z_\text{TC}$ is determined by $\omega$, $\sigma_{xx}$, $C_\text{TG}$, and $C_\text{p1}$. Equation \eqref{eq:overall_impedance} is used in Sec. \ref{sec:sigma_xx_calibration} to calibrate $\sigma_{xx}$.

\section{Equivalent-circuit derivation for the dual-gate TME setup}

We next derive the geometric attenuation factor $\gamma_\text{geo}$ for the dual-gate TME polarization measurement. This is the main attenuation factor of the TME signal and is directly controlled by the compensation method. 

The effective circuit of dual-gate measurement is shown in Fig.~4(b) of the main text, which contains the top-gate geometric capacitance $C_\text{TG}^\text{(geo)}$, the bottom-gate geometric capacitance $C_\text{BG}^\text{(geo)}$, and the sample geometric capacitance $C_\text{S}$ between the top and bottom surfaces. Since $\gamma_\text{geo}$ characterizes how efficiently the TME polarization charge induces an image charge on the gate electrode through the gate dielectric, it is a purely electrostatic capacitive effect and is therefore independent of the quantum capacitance of the sample. However, to avoid introducing excessive definitions, we use $C_\text{TG}$ and $C_\text{BG}$ in Fig.~4(b) to denote $C_\text{TG}^\text{(geo)}$ and $C_\text{BG}^\text{(geo)}$, respectively. To maintain consistency with the analysis in Fig.~4(b), we continue to use $C_\text{TG}$ and $C_\text{BG}$ in this section to refer to geometric capacitances. In other sections of the Supplemental Material, however, $C_\text{TG}$ and $C_\text{BG}$ are defined including the quantum capacitance contribution (Eq.~\eqref{eq:SM_total_capacitance}). Moreover, based on previous studies \cite{inhofer2017, inhofer2018, dartiailh2020}, the quantum capacitance of topological-insulator systems is typically on the order of 10 mF/m$^2$, which is much larger than the typical geometric capacitance per unit area in our experiment, approximately 0.3 mF/m$^2$. This large difference indicates that, even when the quantum capacitance is included, $C_\text{TG}$ and $C_\text{BG}$ remain approximately equal to $C_\text{TG}^\text{(geo)}$ and $C_\text{BG}^\text{(geo)}$, respectively. Therefore, this simplification is well justified.

The intrinsic TME polarization produces opposite surface charges $\pm Q_\text{4D-QHE}$, with
\[
Q_\text{4D-QHE}=A_\text{S}B_\text{AC}\frac{e^2}{2h}N_\text{Ch}^{(2)},
\]
which are considered source charges in the circuit. Let $q_1$ and $q_2$ denote the charges measured in the top and bottom gate lines, respectively.

For the uncompensated case, $q_1$, $q_2$, and $q_\text{S}$ denote the charges on the lower plates (see Fig.~4(b) of the main text) of $C_\text{TG}$, $C_\text{BG}$, and $C_\text{S}$, respectively. Since the top and bottom current preamplifiers are both effectively grounded, the voltages on the top and bottom gates are zero, so the three capacitors form a closed series loop. Charge conservation at the top and bottom sample surfaces gives
\[
Q_\text{4D-QHE} = q_1 - q_\text{S},
\qquad
- Q_\text{4D-QHE} = q_\text{S} - q_2.
\]
The total voltage drop across the three capacitors must be zero:
\begin{equation}
\frac{q_1}{C_\text{TG}}+\frac{q_\text{S}}{C_\text{S}}+\frac{q_2}{C_\text{BG}}=0.
\label{eq:SM_TME_voltage_loop}
\end{equation}
Solving these equations:
\begin{equation}
q_1=q_2=\frac{C_\text{total}}{C_\text{S}}\,Q_\text{4D-QHE} \equiv  \gamma_\text{geo}Q_\text{4D-QHE},
\qquad
C_\text{total}
\equiv
\left(
\frac{1}{C_\text{S}}+\frac{1}{C_\text{TG}}+\frac{1}{C_\text{BG}}
\right)^{-1}.
\label{eq:SM_TME_uncomp}
\end{equation}
The corresponding geometric attenuation factor is
\begin{equation}
\gamma_\text{geo}=\frac{C_\text{total}}{C_\text{S}}.
\label{eq:SM_TME_gamma_uncomp}
\end{equation}

When the compensation is applied, the top and bottom gate capacitances are replaced by the effective capacitances
\begin{equation}
C_\text{TG}^{\rm eff}=\left(\frac{1}{C_\text{TG}}-\frac{1}{C_1}\right)^{-1},
\qquad
C_\text{BG}^{\rm eff}=\left(\frac{1}{C_\text{BG}}-\frac{1}{C_2}\right)^{-1}.
\label{eq:effective_gate_capacitance}
\end{equation}
Therefore, Eq. \eqref{eq:SM_TME_uncomp} remains valid after the substitutions
\[
C_\text{TG}\rightarrow C_\text{TG}^{\rm eff},
\qquad
C_\text{BG}\rightarrow C_\text{BG}^{\rm eff}.
\]
Namely,
\begin{equation}
C_\text{total}^{\rm eff}
=
\left(
\frac{1}{C_\text{S}}+\frac{1}{C_\text{TG}^{\rm eff}}+\frac{1}{C_\text{BG}^{\rm eff}}
\right)^{-1},
\qquad
\gamma_\text{geo}=\frac{C_\text{total}^{\rm eff}}{C_\text{S}},
\qquad
q_1=q_2=\gamma_\text{geo}Q_\text{4D-QHE}.
\label{eq:SM_TME_gamma_eff}
\end{equation}
Thus, the role of the compensation is to increase $C_\text{TG}^{\rm eff}$ and $C_\text{BG}^{\rm eff}$, and hence to increase $\gamma_\text{geo}$ toward unity.

To relate this to the notation used in the main text, we define the compensation ratio
\begin{equation}
\alpha
=
\frac{\frac{1}{C_1}+\frac{1}{C_2}}
{\frac{1}{C_\text{TG}}+\frac{1}{C_\text{BG}}}.
\label{eq:SM_TME_alpha}
\end{equation}
Then
\[
\frac{1}{C_\text{TG}^{\rm eff}}+\frac{1}{C_\text{BG}^{\rm eff}}
=
(1-\alpha)\left(\frac{1}{C_\text{TG}}+\frac{1}{C_\text{BG}}\right),
\]
so that
\begin{equation}
\gamma_\text{geo}
=
\frac{1}{
1+C_\text{S}(1-\alpha)\left(\frac{1}{C_\text{TG}}+\frac{1}{C_\text{BG}}\right)
}.
\label{eq:SM_TME_gamma_alpha}
\end{equation}

Equations \eqref{eq:SM_TME_uncomp}--\eqref{eq:SM_TME_gamma_alpha} are the key results of the dual-gate TME circuit analysis. They show that $\gamma_\text{geo}$ is the dominant electrostatic attenuation factor of the TME measurement and that it is directly tuned by the compensation through the effective gate capacitances.

\section{Dissipation time scale analysis}

The dissipation time scales can be understood from a simple RC-decay picture. In both the QAH and TME measurements, the relaxation is governed by the longitudinal conduction of the sample, so the effective resistance scales as
\[
R\sim \frac{1}{\sigma_{xx}}.
\]

For the compensated QAH charge accumulation, the sample voltage is approximately
\[
U_\text{QAH} \approx Q_\text{a}\left(\frac{1}{C_\text{TG}}-\frac{1}{C_1}\right)=\frac{Q_\text{a}}{C_\text{TG}^{\rm eff}}.
\]
The corresponding time scale is
\begin{equation}
\tau_\text{QAH}\sim R\,C_\text{TG}^{\rm eff}\sim \frac{C_\text{TG}^{\rm eff}}{\sigma_{xx}}.
\label{eq:SM_tau_QAH}
\end{equation}
For symmetric single-gate compensation, $C_\text{TG}^{\rm eff}=C_\text{TG}/(1-\alpha)$, so equivalently
\[
\tau_\text{QAH}\sim \frac{C_\text{TG}}{\sigma_{xx}(1-\alpha)}.
\]

For the TME setup, the measurable gate charge is reduced by the geometric factor,
\[
q_1=q_2=\gamma_\text{geo}Q_\text{4D-QHE}.
\]
The corresponding surface voltage is
\[
U_\text{TME}=\frac{q_1}{C_\text{TG}^{\rm eff}}
=\frac{\gamma_\text{geo}Q_\text{4D-QHE}}{C_\text{TG}^{\rm eff}}.
\]
Comparing this with the QAH voltage scale $U_\text{QAH}$ under the same $\sigma_{xx}$,
\begin{equation}
\frac{\tau_\text{TME}}{\tau_\text{QAH}} = \frac{1}{\gamma_\text{geo}}.
\label{eq:SM_tau_ratio}
\end{equation}
Therefore,
\begin{equation}
\tau_\text{TME}
\sim
\frac{\tau_\text{QAH}}{\gamma_\text{geo}}
=
\frac{C_\text{TG}^{\rm eff}}{\sigma_{xx}}
\frac{C_\text{S}}{C_\text{total}^{\rm eff}}.
\label{eq:SM_tau_TME}
\end{equation}

In summary,
\begin{equation}
\tau_\text{QAH}\sim \frac{C_\text{TG}^{\rm eff}}{\sigma_{xx}},
\qquad
\tau_\text{TME}\sim \frac{C_\text{TG}^{\rm eff}}{\sigma_{xx}}\frac{C_\text{S}}{C_\text{total}^{\rm eff}}
=\frac{\tau_\text{QAH}}{\gamma_\text{geo}}.
\label{eq:SM_tau_summary}
\end{equation}

\section{Transport properties of the QAH sample}

This section summarizes the transport characterization used to identify the QAH regime and the gate-voltage working point.

\subsection{Hall-bar transport and gate voltage dependence}

\begin{figure}[htbp]
  \centering
  \includegraphics[width=1\linewidth]{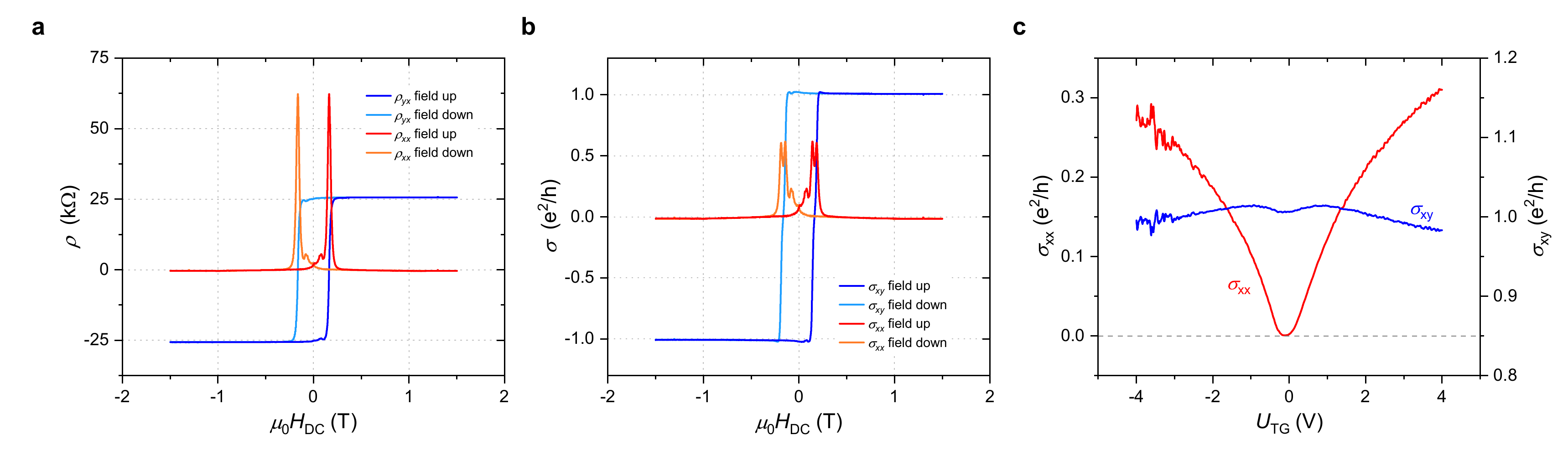}
  \caption{\textbf{Transport characterization of the Hall-bar sample.}
  (a) DC magnetic-field dependence of $\rho_{xx}$ and $\rho_{yx}$ measured at $U_\text{TG}=0$ V and $T\approx 50$ mK.
  (b) Corresponding $\sigma_{xx}$ and $\sigma_{xy}$ obtained from the data in (a).
  (c) $U_\text{TG}$ dependence of $\sigma_{xx}$ and $\sigma_{xy}$ measured at $|\mu_0 H_\text{DC}|=1.5$ T.}
  \label{fig:figS_transport_hallbar}
\end{figure}

The Hall-bar sample was characterized by standard low-frequency lock-in measurements with an excitation current of 10 nA. Figure~\ref{fig:figS_transport_hallbar}(a) shows the longitudinal and Hall resistivities, $\rho_{xx}$ and $\rho_{yx}$, measured at $U_\text{TG}=0$ V and $T\approx 50$ mK. The corresponding conductivities, obtained from tensor inversion, are shown in Fig.~\ref{fig:figS_transport_hallbar}(b). The sample exhibits a robust QAH state, with quantized $\sigma_{xy}$ and strongly suppressed $\sigma_{xx}$ outside the coercive-field region.

To determine the optimal gate voltage, we measured $\sigma_{xx}$ and $\sigma_{xy}$ as a function of $U_\text{TG}$ at $\mu_0 H_\text{DC}=\pm1.5$ T. As shown in Fig.~\ref{fig:figS_transport_hallbar}(c), $\sigma_{xy}$ is quantized and $\sigma_{xx}$ reaches its minimum near $U_\text{TG}=0$ V. Therefore, no DC gate voltage was applied during the charge-accumulation measurements in the main text.

\subsection{Determination of the small $\sigma_{xx}$ from the Corbino disk-shaped sample}

\begin{figure}[htbp]
  \centering
  \includegraphics[width=1\linewidth]{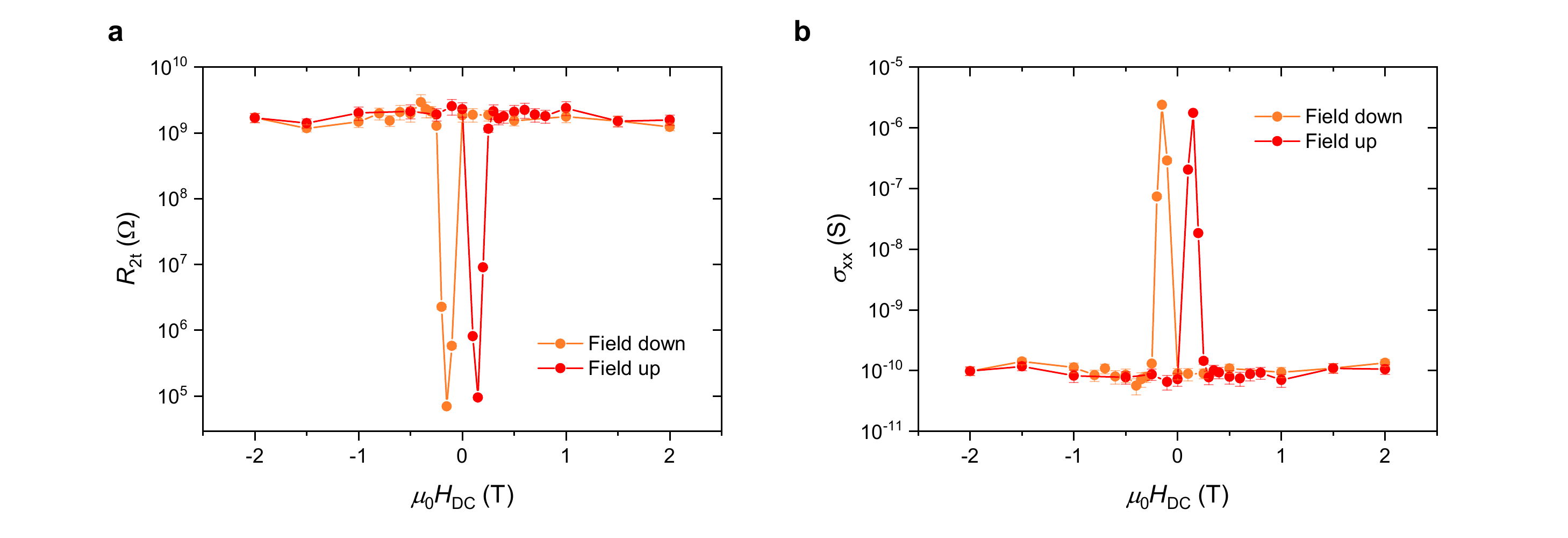}
  \caption{\textbf{Two-terminal transport of the Corbino disk-shaped sample.}
  (a) Two-terminal resistance $R_\text{2t}$ measured between the inner and outer contacts as a function of the DC magnetic field at base temperature.
  (b) $\sigma_{xx}$ calculated from Eq. \eqref{eq:SM_Corbino_sigma}. Error bars represent the statistical uncertainty.}
  \label{fig:figS_transport_corbino}
\end{figure}

To estimate the dissipation level more accurately in the disk-shaped devices, we performed a two-terminal measurement on the Corbino disk-shaped sample. An AC voltage $U_\text{2t}=\SI{40}{\micro\volt}$ at $f=7.167$ Hz was applied between the inner and outer contacts, and the current $I_\text{2t}$ was measured using a transimpedance current preamplifier. Figure~\ref{fig:figS_transport_corbino}(a) shows the resulting two-terminal resistance $R_\text{2t}=U_\text{2t}/I_\text{2t}$ as a function of the DC magnetic field at base temperature.

The longitudinal conductance was obtained from the Corbino relation
\begin{equation}
\sigma_{xx}=\frac{\ln(r_2/r_1)}{2\pi}\frac{1}{R_\text{2t}},
\label{eq:SM_Corbino_sigma}
\end{equation}
where $r_1=\SI{350}{\micro\meter}$ and $r_2=\SI{1000}{\micro\meter}$ are the inner and outer radii. The resulting $\sigma_{xx}$ is shown in Fig.~\ref{fig:figS_transport_corbino}(b). In the QAH plateau region, $\sigma_{xx}$ is suppressed to the order of $10^{-10}$ S, providing the relevant scale for the low-dissipation charge-accumulation measurements.

\section{Calibration of the AC magnetic field}
\label{sec:coil_calibration}

The charge-accumulation measurements require a calibrated AC magnetic field $B_\text{AC}$. Because the relation between the coil excitation voltage $U_\text{coil}$ and the generated field is frequency dependent, the calibration was performed in situ using a Hall sensor placed near the sample. We first define and calibrate the complex proportionality coefficient
\begin{equation}
    k_\text{coil}(f)=\frac{B_\text{AC}}{U_\text{coil}},
\end{equation}
so that the AC magnetic field used in the charge measurements is obtained from the applied excitation voltage through $B_\text{AC}=k_\text{coil}U_\text{coil}$.

\begin{figure}[htbp]
  \centering
  \includegraphics[width=1\linewidth]{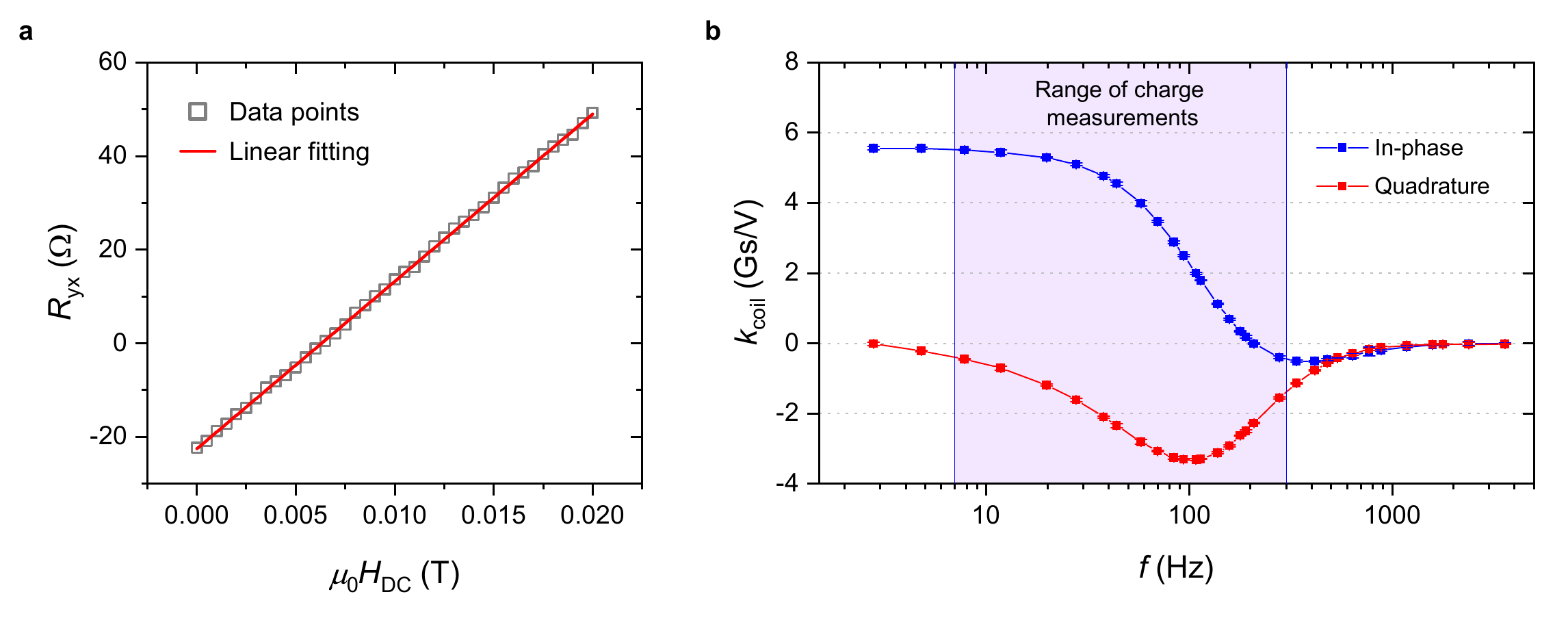}
  \caption{\textbf{Calibration of the AC magnetic field generated by the coil.}
  (a) DC-field dependence of the Hall resistance of the Hall sensor and linear fitting of $k_\text{Hall}$.
  (b) Frequency dependence of the calibrated $k_\text{coil}$, the AC magnetic field per unit source voltage. The in-phase and quadrature components are plotted separately.}
  \label{fig:figS_Bac}
\end{figure}

To calibrate $k_\text{coil}$, a DC current of $I_\text{sensor} = \SI{1}{\micro\ampere}$ was applied to the Hall sensor under a small DC magnetic field of $0.01$ T, and the AC Hall voltage ($U_\text{Hall}$) was measured by a lock-in amplifier referenced to the source voltage of the AC coil. The AC magnetic field was then obtained from
\begin{equation}
B_\text{AC}=\frac{U_\text{Hall}}{I_\text{sensor}k_\text{Hall}},
\label{eq:SM_Bac_calibration}
\end{equation}
where $k_\text{Hall}$ is the Hall coefficient obtained from a separate DC-field calibration. Figure~\ref{fig:figS_Bac}(a) shows the calibration of $k_\text{Hall}$, and the fitting results are summarized in Table~\ref{tab:hall_fit}. Figure~\ref{fig:figS_Bac}(b) shows the frequency dependence of the calibrated $k_\text{coil}$.

\begin{figure}[htbp]
  \centering
  \includegraphics[width=0.9\linewidth]{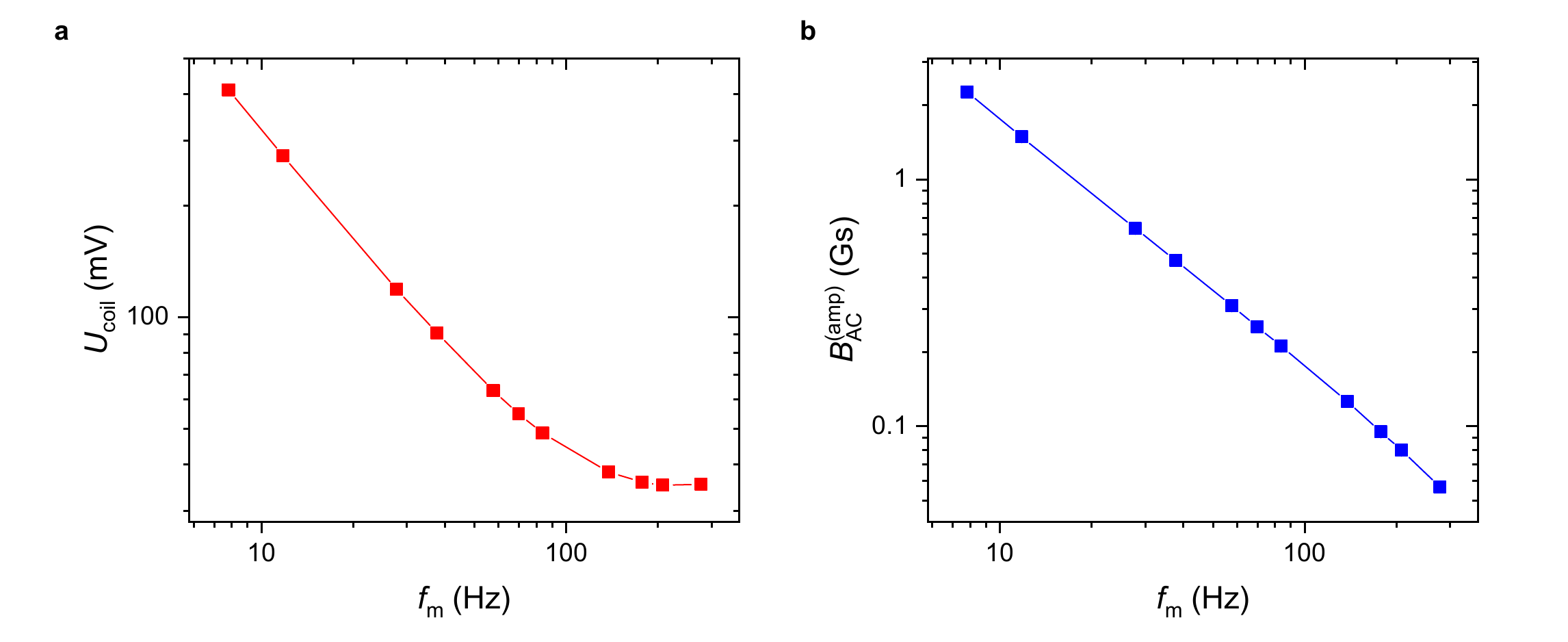}
  \caption{\textbf{Calibration of $B_\text{AC}$ for constant $\sigma_{xx}$.}
  (a) Calibrated $U_\text{coil}$ as a function of the AC-field frequency $f_\text{m}$.
  (b) Corresponding $B_\text{AC}$ amplitude as a function of $f_\text{m}$.}
  \label{fig:figS_excitation_voltage}
\end{figure}

In the compensated QAH charge-accumulation measurements, the sample temperature, and therefore $\sigma_{xx}$, was controlled by the eddy-current heating generated by the AC magnetic field. To keep $\sigma_{xx}$ constant during the frequency sweep, we adjusted $U_\text{coil}$ at each measurement frequency $f_\text{m}$ (magnetic field frequency) so that the impedance between the top gate and the grounded contact, $Z_\text{TC}$, measured at $f_\text{c}=117.777$ Hz (calibration frequency), remained at the same target value. Figure~\ref{fig:figS_excitation_voltage}(a) shows the calibrated $U_\text{coil}(f_\text{m})$, and Fig.~\ref{fig:figS_excitation_voltage}(b) shows the corresponding $B_\text{AC}$ obtained from $k_\text{coil}(f_\text{m})$.

After this calibration, the corresponding dissipation levels were determined independently from the $f_\text{c}$ dependence of $Z_\text{TC}$, as described in Sec. \ref{sec:sigma_xx_calibration}. The resulting constant-$\sigma_{xx}$ conditions correspond to $\sigma_{xx}=(2.35\pm0.01)\times10^{-8}$ S at $\mu_0H_\text{DC}=0.5$ T and $\sigma_{xx}=(4.08\pm0.09)\times10^{-8}$ S at $\mu_0H_\text{DC}=0$ T.

\begin{table}[htbp]
\centering
\caption{Linear fitting results of $k_\text{Hall}$ near a DC field of 0.01 T.}
\label{tab:hall_fit}
\setlength{\tabcolsep}{10pt}
\begin{tabular}{cc}
\toprule
Parameter & Value \\
\midrule
Intercept $R_0$ ($\Omega$) & $-22.6 \pm 0.1$ \\
Slope $k_\text{Hall}$ ($\Omega/\text{T}$) & $3577 \pm 9$ \\
Fitting $R^2$ (COD) & $0.99976$ \\
\bottomrule
\end{tabular}
\end{table}

\section{Calibration of $\sigma_{xx}$ in simple disk-shaped sample}
\label{sec:sigma_xx_calibration}

In this section, we calibrate $\sigma_{xx}$ for the simple disk-shaped sample by fitting the frequency dependence of the gate-contact impedance $Z_\text{TC}$ to Eq. \eqref{eq:overall_impedance}. The fit sets $\sigma_{xx}$, $C_\text{TG}$, and $C_\text{p1}$ as fitting parameters.

As described in Sec. \ref{sec:coil_calibration}, the AC magnetic field was first calibrated so that $Z_\text{TC}$ at a fixed frequency $f_\text{c}=\SI{117.777}{\hertz}$ remained constant during the $f_\text{m}$ sweep. The corresponding $\sigma_{xx}$ was then obtained from the full $f_\text{c}$ dependence of $Z_\text{TC}$ at $f_\text{m}=\SI{83.7777}{\hertz}$.

\begin{figure}[htbp]
  \centering
  \includegraphics[width=1\linewidth]{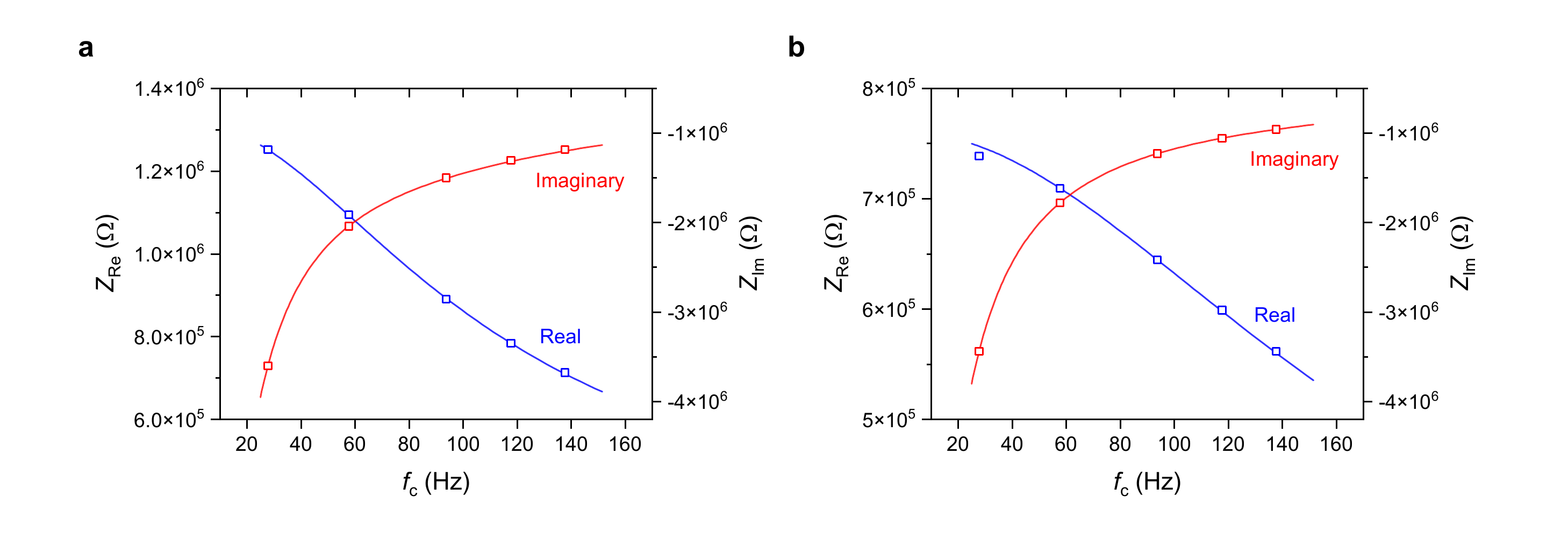}
  \caption{\textbf{Calibration of $\sigma_{xx}$ from the gate-contact impedance of the simple disk-shaped sample.}
  (a) Frequency dependence of the real and imaginary parts of the measured impedance at $\mu_0H_\text{DC}=0.5$ T. Symbols are experimental data and solid lines are fits to the impedance model.
  (b) Same as in (a), but measured at $\mu_0H_\text{DC}=0$ T.}
  \label{fig:figS_impedance_fit}
\end{figure}

The fitting results are summarized in Table~\ref{tab:SM_impedance_fit}. Figure~\ref{fig:figS_impedance_fit}(a) shows the measured real and imaginary parts of $Z_\text{TC}$ at $\mu_0H_\text{DC}=0.5$ T, together with the fitting curves. Figure~\ref{fig:figS_impedance_fit}(b) shows the corresponding data at $\mu_0H_\text{DC}=0$ T. In both cases, the fitted curves reproduce the measured response well.

The extracted $\sigma_{xx}$ values are used in the analysis of the compensated QAH charge-accumulation measurements. The fitted $C_\text{TG}$ and $C_\text{p1}$ are consistent with the independent capacitance calibration given in the next section.

\begin{table}[htbp]
\centering
\caption{Fitting results of the gate-contact impedance measurement for the simple disk-shaped sample.}
\label{tab:SM_impedance_fit}
\setlength{\tabcolsep}{10pt}
\begin{tabular}{cccc}
\toprule
$\mu_0H_\text{DC}$ & $\sigma_{xx}$ (S) & $C_\text{TG}$ (nF) & $C_\text{p1}$ (nF) \\
\midrule
0.5 T & $(2.35 \pm 0.01)\times10^{-8}$ & $1.506 \pm 0.001$ & $0.202 \pm 0.002$ \\
0 T   & $(4.08 \pm 0.09)\times10^{-8}$ & $1.508 \pm 0.009$ & $0.200 \pm 0.010$ \\
\bottomrule
\end{tabular}
\end{table}

\section{Capacitance calibration of the compensation system}

This section summarizes the capacitance calibration relevant to the compensation method. For the single-gate QAH setup, the relevant capacitances are the top-gate capacitance $C_\text{TG}$ (including the quantum-capacitance contribution), the overlap capacitance $C_\text{p1}$ between the top gate and the contact, and the parasitic capacitance $C_\text{p0}$ of the gate line. For the dual-gate TME setup, the relevant capacitances are $C_\text{TG}$, $C_\text{BG}$, the parasitic capacitances of the two gate lines, and the sample geometric capacitance $C_\text{S}$.

\subsection{Capacitance calibration for the single-gate QAH setup}

For the single-gate QAH setup, the calibration was performed near the coercive field, $\mu_0H_\text{DC}=\SI{0.15}{\tesla}$, where the sample resistance is much smaller than the capacitive impedance. The applied AC voltage amplitude was fixed at $U_\text{c}=\SI{100}{\micro\volt}$.

\begin{figure}[htbp]
  \centering
  \includegraphics[width=1\linewidth]{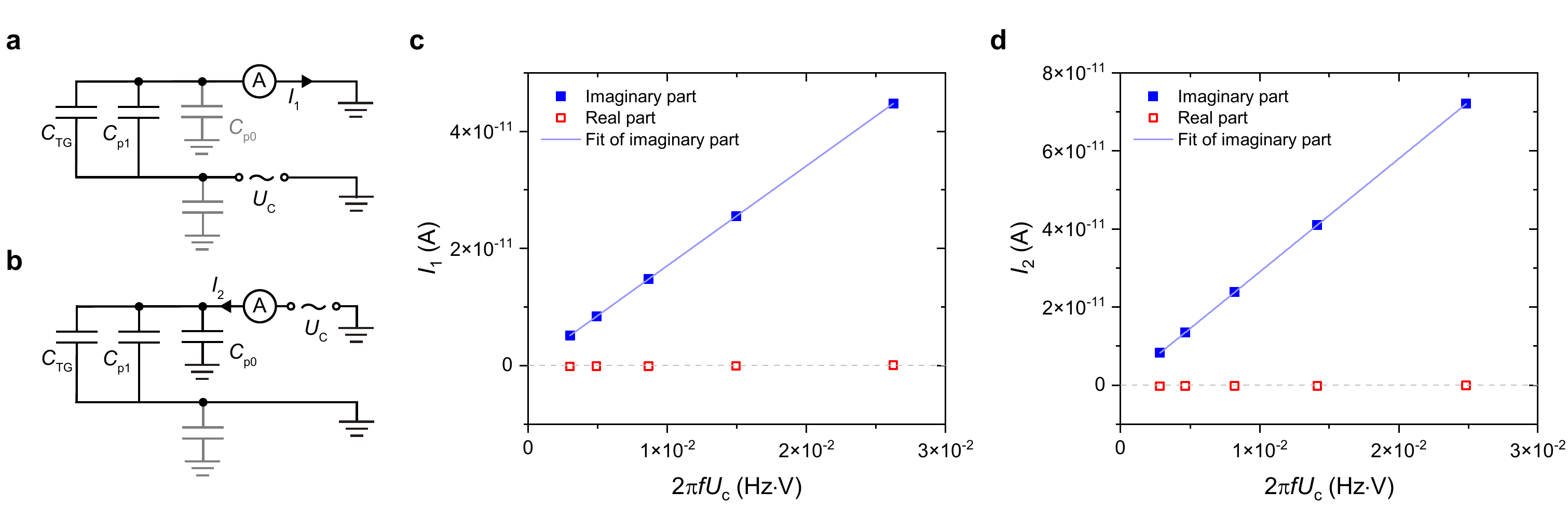}
  \caption{\textbf{Capacitance calibration for the single-gate QAH setup.}
  (a,b) Schematics of the two capacitance measurement configurations. In both panels, the gray part denotes the circuit branch that does not contribute to the measured current.
  (c,d) Real and imaginary parts of the measured current as a function of $2\pi f U_\text{c}$ for the configuration in (a) and in (b). Solid lines are linear fits of the imaginary part to Eq. \eqref{eq:SM_I1_cap} and Eq. \eqref{eq:SM_I2_cap}, respectively.}
  \label{fig:figS_cap_QAH}
\end{figure}

Two configurations were used for the calibration, as shown in Fig.~\ref{fig:figS_cap_QAH}(a) and \ref{fig:figS_cap_QAH}(b). In both cases the current was measured in the top-gate line. In the first configuration, the measured current contains only the contributions from $C_\text{TG}$ and $C_\text{p1}$,
\begin{equation}
I_1=i\,2\pi f\,(C_\text{TG}+C_\text{p1})\,U_\text{c}.
\label{eq:SM_I1_cap}
\end{equation}
In the second configuration, the current additionally includes $C_\text{p0}$,
\begin{equation}
I_2=i\,2\pi f\,(C_\text{TG}+C_\text{p1}+C_\text{p0})\,U_\text{c}.
\label{eq:SM_I2_cap}
\end{equation}
Therefore, linear fits of the imaginary part of the current versus $2\pi f U_\text{c}$ directly give $(C_\text{TG}+C_\text{p1})$ and $(C_\text{TG}+C_\text{p1}+C_\text{p0})$.

The corresponding data are shown in Fig.~\ref{fig:figS_cap_QAH}(c) and \ref{fig:figS_cap_QAH}(d), and the extracted values are summarized in Table~\ref{tab:SM_cap_QAH}. To separate $C_\text{TG}$ and $C_\text{p1}$, we use the device geometry. The sample area is $A_\text{S}=\SI{4.52}{\milli\meter\squared}$ and the top gate area is $A_\text{TG}=\SI{4.94}{\milli\meter\squared}$, so the top gate slightly overlaps the contact region. Assuming the same dielectric thickness and dielectric constant in both regions and a large quantum capacitance, the capacitance ratio is approximated by the area ratio,
\begin{equation}
\frac{C_\text{TG}}{C_\text{p1}}
\approx
\frac{A_\text{S}}{A_\text{TG}-A_\text{S}}.
\label{eq:SM_area_ratio}
\end{equation}
This gives $C_\text{TG}$ and $C_\text{p1}$ separately, while $C_\text{p0}$ follows from the difference of the two fitted totals. The calibrated capacitances are sufficient to control the compensation ratio $\alpha$ at the \SI{1}{\percent} level.

\begin{table}[htbp]
\centering
\caption{Capacitance calibration results for the single-gate QAH setup. The two quantities on the left are obtained directly from linear fitting of the imaginary part of current versus $2\pi f U_\text{c}$. The capacitances on the right are further extracted using the geometric area ratio.}
\label{tab:SM_cap_QAH}
\setlength{\tabcolsep}{8pt}
\begin{tabular}{cc|ccc}
\toprule
$C_\text{TG}+C_\text{p1}$ (nF) & $C_\text{TG}+C_\text{p1}+C_\text{p0}$ (nF) & $C_\text{TG}$ (nF) & $C_\text{p1}$ (nF) & $C_\text{p0}$ (nF) \\
\midrule
$1.7043 \pm 0.0004$ & $2.8994 \pm 0.0007$ & $1.559 \pm 0.001$ & $0.145 \pm 0.001$ & $1.195 \pm 0.001$ \\
\bottomrule
\end{tabular}
\end{table}

\subsection{Capacitance calibration for the dual-gate TME setup}

For the dual-gate TME setup, the relevant capacitances are $C_\text{TG}$, $C_\text{BG}$, the parasitic capacitances of the two gate lines, and the sample geometric capacitance $C_\text{S}$. The overlap capacitance analogous to $C_\text{p1}$ can be estimated from the device geometry and is neglected here.

The gate capacitances $C_\text{TG}$ and $C_\text{BG}$, together with the corresponding parasitic capacitances, can be calibrated in the same way as in the single-gate case by tuning the sample surface to the coercive-field region and fitting the linear frequency dependence of the imaginary current.

The sample geometric capacitance $C_\text{S}$ is calibrated separately in the axion-insulator state, where the sample is insulating. An AC voltage is applied to one gate and the current is measured in the other gate line, so that the line parasitic capacitances do not contribute. The measured total capacitance ($C_\text{total}$) then satisfies
\begin{equation}
C_\text{total}^{-1}=(C_\text{TG}^{(\text{geo})})^{-1}+(C_\text{BG}^{(\text{geo})})^{-1}+C_\text{S}^{-1} \approx C_\text{TG}^{-1}+C_\text{BG}^{-1}+C_\text{S}^{-1},
\label{eq:SM_Cs_calibration}
\end{equation}
from which $C_\text{S}$ is obtained.

\section{Additional data of QAH charge accumulation compensation}

In this section, we provide additional compensation data beyond those shown in the main text. These data further verify that the active capacitive compensation quantitatively suppresses the dissipation-induced attenuation of the QAH charge-accumulation signal in different sample geometries and attenuation conditions.

\subsection{Simple disk-shaped sample at $\mu_0H_\text{DC}=0$ T}

\begin{figure}[htbp]
  \centering
  \includegraphics[width=1\linewidth]{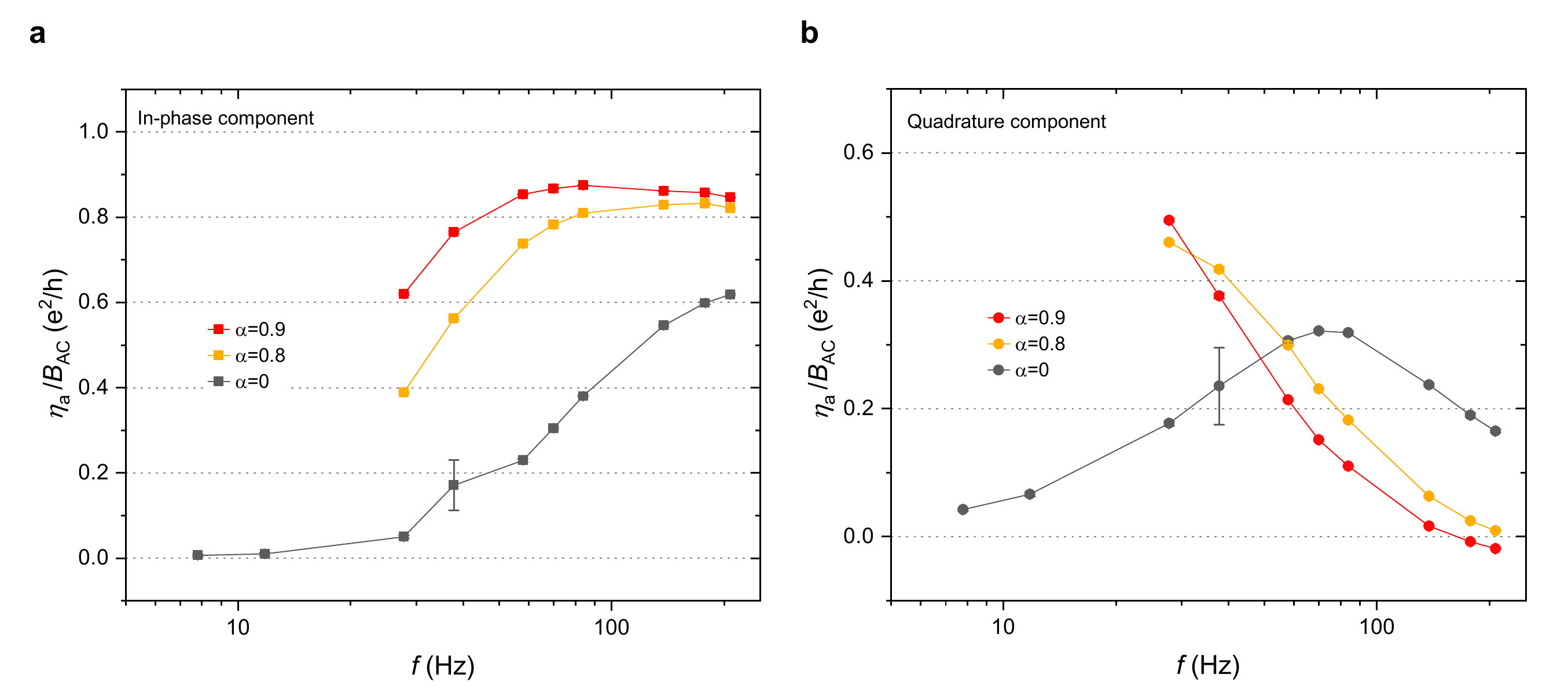}
  \caption{\textbf{Additional compensated charge accumulation data for the simple disk-shaped sample at $\mu_0H_\text{DC}=0$ T.}
  Frequency dependence of the QAH charge-accumulation signal measured for $\alpha=0$, $0.8$, and $0.9$. (a) In-phase component. (b) Quadrature component.}
  \label{fig:figS_QAH_simpledisk_0T}
\end{figure}

Fig.~\ref{fig:figS_QAH_simpledisk_0T} shows the frequency dependence of the QAH charge-accumulation signal measured on the simple disk-shaped sample at $\mu_0H_\text{DC}=0$ T, where $\sigma_{xx}=4.08\times10^{-8}$ S was kept unchanged. The measurements were performed for several compensation ratios, $\alpha=0$, $0.8$, and $0.9$. The in-phase and quadrature components are plotted separately, in the same manner as in Fig.~2(a) and (b) of the main text.

The data show the same systematic trend as that observed at $\mu_0H_\text{DC}=0.5$ T in the main text: increasing $\alpha$ enhances the in-phase component toward the quantized value and suppresses the quadrature component, consistent with the theoretical dissipation model.

\subsection{Corbino disk-shaped sample at $\mu_0H_\text{DC}=0$ T}

\begin{figure}[htbp]
  \centering
  \includegraphics[width=1\linewidth]{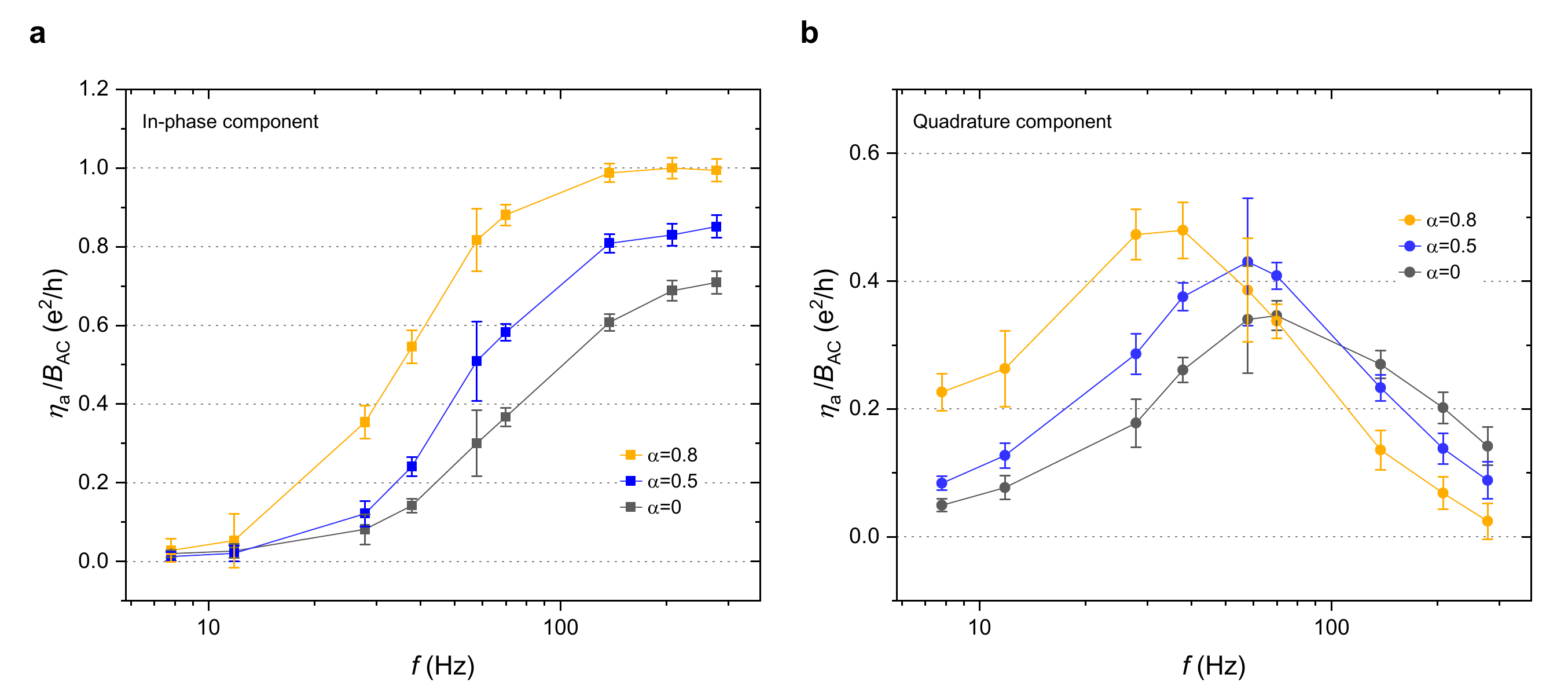}
  \caption{\textbf{Additional compensated charge accumulation data on the Corbino disk-shaped sample at $\mu_0H_\text{DC}=0$ T.}
  Frequency dependence of the QAH charge-accumulation signal measured for $\alpha=0$, $0.5$, and $0.8$. (a) In-phase component. (b) Quadrature component.}
  \label{fig:figS_QAH_corbino_0T}
\end{figure}

Fig.~\ref{fig:figS_QAH_corbino_0T} shows the frequency dependence of the QAH charge-accumulation signal measured on the Corbino disk-shaped sample at $\mu_0H_\text{DC}=0$ T. The measurements were performed for $\alpha=0$, $0.5$, and $0.8$. $\sigma_{xx}$ was kept unchanged ($5.01\times 10^{-9}$ S) during the frequency sweep. The top gate of this sample has an inner radius of \SI{450}{\micro\meter} and an outer radius of \SI{900}{\micro\meter}, and does not completely cover the sample area (with inner and outer radii of \SI{350}{\micro\meter} and \SI{1000}{\micro\meter} respectively). However, using the top-gate area as the effective area still yields compensated signals close to quantization, consistent with Fig.~2(a) and (b) of the main text.

\section{Treatment of parasitic capacitance in the compensation circuit}

The discussion in the main text assumes an ideal circuit without parasitic capacitance. In the real measurement, however, the parasitic capacitances modify both the feedback condition and the measured charge signal. Therefore, the feedback parameter $C_1$ must be corrected, and the measured signals must be converted back to the true accumulated charge.

\begin{figure}[htbp]
  \centering
  \includegraphics[width=0.4\linewidth]{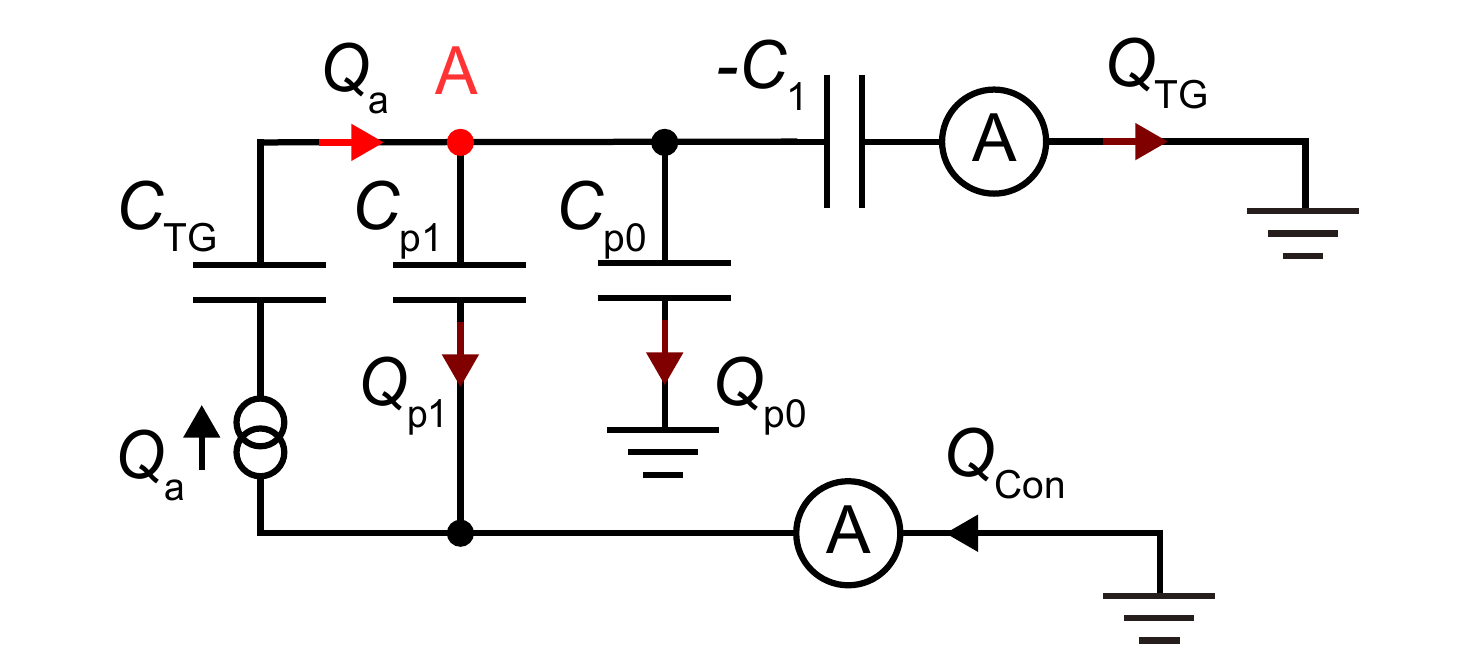}
  \caption{\textbf{Equivalent circuit of the compensated measurement in the presence of parasitic capacitances $C_\text{p1}$ and $C_\text{p0}$.}}
  \label{fig:figS_parasitic_comp_circuit}
\end{figure}

For the single-gate QAH setup, the relevant parasitic capacitances $C_\text{p1}$ and $C_\text{p0}$ have been calibrated in Table~\ref{tab:SM_cap_QAH}. The corresponding circuit is shown in Fig.~\ref{fig:figS_parasitic_comp_circuit}. To realize a target compensation ratio $\alpha$, the gate voltage (node A in Fig.~\ref{fig:figS_parasitic_comp_circuit}) should satisfy
\begin{equation}
U_\text{TG}=-\alpha \frac{Q_\text{a}}{C_\text{TG}}.
\label{eq:SM_parasitic_UTG_target}
\end{equation}
Here $C_\text{TG}$ includes the quantum capacitance contribution as described in the dissipative charge-accumulation model. At the same time, node A is connected to the three grounded branches $C_\text{p1}$, $C_\text{p0}$, and $-C_1$, so its potential is
\begin{equation}
U_\text{TG}=\frac{Q_\text{a}}{C_\text{p1}+C_\text{p0}-C_1}.
\label{eq:SM_parasitic_UTG_actual}
\end{equation}
Equating the two expressions gives the corrected feedback capacitance
\begin{equation}
C_1=\frac{C_\text{TG}}{\alpha}+C_\text{p0}+C_\text{p1}.
\label{eq:SM_parasitic_C1}
\end{equation}

Using this result, the true accumulated charge can be obtained from the top-gate signal $Q_\text{TG}$ or the contact signal $Q_\text{Con}$ as
\begin{equation}
Q_\text{a}
=
\frac{Q_\text{TG}}{1+\alpha(C_\text{p0}+C_\text{p1})/C_\text{TG}},
\label{eq:SM_parasitic_Qa_from_QTG}
\end{equation}
\begin{equation}
Q_\text{a}
=
\frac{Q_\text{Con}}{1+\alpha C_\text{p1}/C_\text{TG}}.
\label{eq:SM_parasitic_Qa_from_QCon}
\end{equation}
Eliminating $\alpha$ between these two expressions gives an $\alpha$-independent form,
\begin{equation}
Q_\text{a}
=
Q_\text{Con}\left(1+\frac{C_\text{p1}}{C_\text{p0}}\right)
-
Q_\text{TG}\frac{C_\text{p1}}{C_\text{p0}},
\label{eq:SM_parasitic_Qa_general}
\end{equation}
which is used in the charge data analysis of the main text.

\begin{figure}[htbp]
  \centering
  \includegraphics[width=1\linewidth]{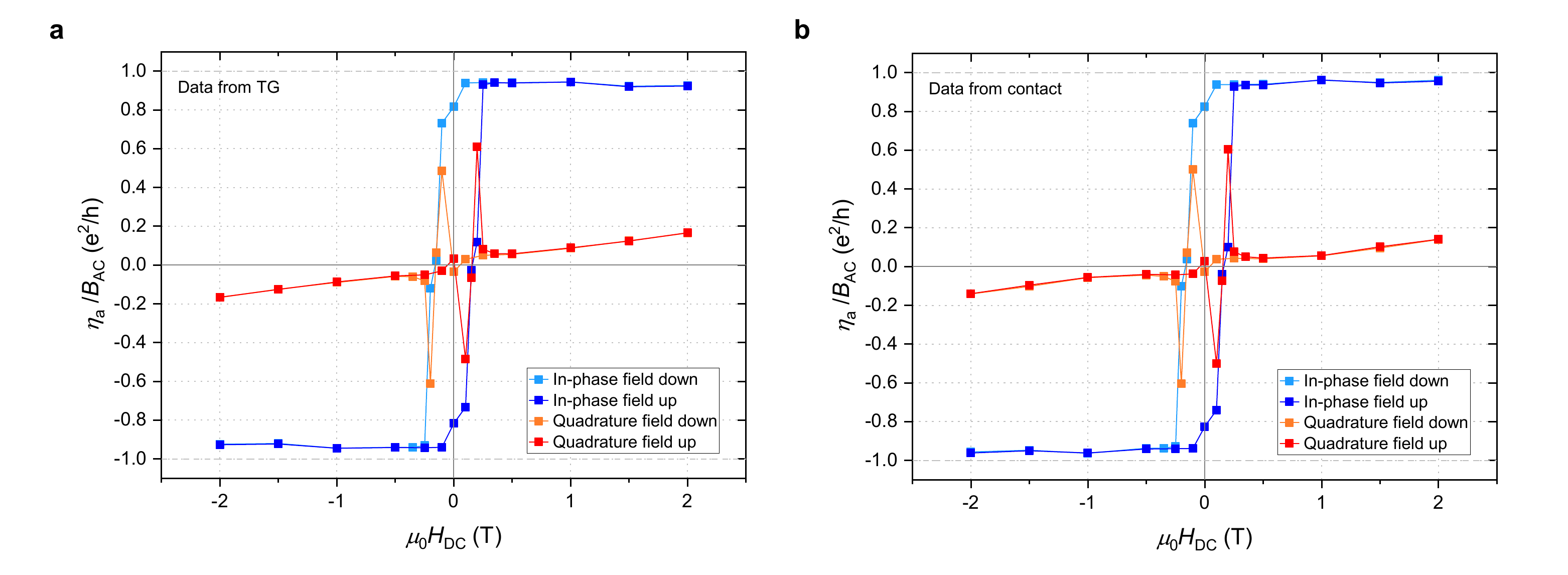}
  \caption{\textbf{Compensated QAH charge-accumulation hysteresis loops calculated from the top-gate data and the contact data.}
  (a,b) Compensated field-sweep data at $f=277.777$ Hz and $\alpha=0.9$, processed from the top-gate signal using Eq. \eqref{eq:SM_parasitic_Qa_from_QTG} (a) and from the contact signal using Eq. \eqref{eq:SM_parasitic_Qa_from_QCon} (b).}
  \label{fig:figS_parasitic_comp}
\end{figure}

Since future dual-gate TME measurements will rely only on the gate lines for signal readout, it is important to verify that the top-gate signal alone still gives the correct result after the parasitic-capacitance correction. Figure~\ref{fig:figS_parasitic_comp}(a) and (b) show the compensated field-sweep data extracted from the top-gate signal using Eq. \eqref{eq:SM_parasitic_Qa_from_QTG} and from the contact signal using Eq. \eqref{eq:SM_parasitic_Qa_from_QCon}, respectively. The data were measured at $f=277.777$ Hz with $\alpha=0.9$. The two results agree with Fig.~3(b) of the main text, confirming that the parasitic-capacitance calibration and the corresponding data treatment are reliable.

\clearpage

\bibliography{Supplemental_material}